\documentclass[journal]{IEEEtran}
\usepackage[T1]{fontenc}% optional T1 font encoding

\usepackage{cite}
\usepackage[colorlinks,linkcolor=black,citecolor=black, urlcolor=black]{hyperref}
\usepackage{soul}

\usepackage{graphicx}
\usepackage{amssymb}
\usepackage{amsmath}
\usepackage{amsthm}
\usepackage{bm}

\usepackage[linesnumbered,ruled]{algorithm2e}
\usepackage{booktabs}
\usepackage{subfigure}

\begin{document}

% \title{Feedback-Free Resource Scheduling: Towards Flexible Multi-BS Cooperation in FD-RAN}

\title{Feedback-Free Resource Scheduling for Flexible Multi-BS Cooperation in FD-RAN
}

\author{Jingbo~Liu,
Jiacheng~Chen,~\IEEEmembership{Member,~IEEE,}
Zeyu~Sun,~\IEEEmembership{Student~Member,~IEEE,}
Bo~Qian,~\IEEEmembership{Member,~IEEE,}
and~Haibo~Zhou,~\IEEEmembership{Senior~Member,~IEEE}

\IEEEcompsocitemizethanks
{
\IEEEcompsocthanksitem J. Liu is with the School of Electronic Information and Electrical Engineering, Shanghai Jiao Tong University, Shanghai 200240, China, and also with the Pengcheng Laboratory, Shenzhen 518000, China
(e-mail: liujingbo@sjtu.edu.cn).
\IEEEcompsocthanksitem J. Chen is with the Department of Strategic and Advanced Interdisciplinary Research, Pengcheng Laboratory, Shenzhen 518000, China
(e-mail: chenjch02@pcl.ac.cn).
\IEEEcompsocthanksitem Bo Qian is with the Information Systems Architecture Science Research Division, National Institute of Informatics, Tokyo 101-8430, Japan (e-mail: boqian@ieee.org).
% E-mail: zongxiliu@smail.nju.edu.cn, haibozhou@nju.edu.cn.
\IEEEcompsocthanksitem Z. Sun and H. Zhou are with the School of Electronic Science and Engineering, Nanjing University, Nanjing 210023, China. 
Z. Sun is also with the Pengcheng Laboratory, Shenzhen 518000, China
(e-mail: zeyusun@smail.nju.edu.cn; haibozhou@nju.edu.cn).
}
\thanks
{
% the National Natural Science Foundation Original Exploration Project of China under Grant 62250004,
This work was supported in part by  the Natural Science Foundation of China (NSFC) under Grant 62271244, the Natural Science Fund for Distinguished Young Scholars of Jiangsu Province under Grant BK20220067, the High-level Innovation and Entrepreneurship Talent Introduction Program Team of Jiangsu Province under Grant JSSCTD202202, and the Major Key Project of PCL (PCL2024A03).
(Corresponding authors: Jiacheng Chen; Haibo Zhou.)
}
}

\maketitle

\begin{abstract}
Flexible cooperation among base stations (BSs) is critical to improve resource utilization efficiency and meet personalized user demands. 
However, its practical implementation is hindered by the current radio access network (RAN), which relies on the coupling of uplink and downlink transmissions and channel state information feedback with inherent issues such as overheads and delays. 
To overcome these limitations, we consider the fully-decoupled RAN (FD-RAN), in which uplink and downlink functionalities are independent, and feedback-free MIMO transmission is adopted at the physical layer.
To further deliver flexible cooperation in FD-RAN, we study feedback-free downlink multi-BS resource scheduling under the practical scheduling process.
The problem is considered based on network load conditions. 
In heavy-load states where it is impossible for all user demands to be met, an optimal greedy algorithm is proposed, maximizing the weighted sum of user demand satisfaction rates.
In light-load states where at least one solution exists to satisfy all user demands, an optimal two-stage resource allocation algorithm is designed to further minimize network energy consumption by leveraging the flexibility of cooperation.
Extensive simulations validate the superiority of proposed algorithms in performance and running time, and highlight the potential for realizing flexible cooperation in practice.

\end{abstract}


\section{Introduction} \label{intro}
\IEEEPARstart{C}{ooperation} among multiple base stations (BSs), which enhances coverage and throughput through joint transmission (JT) and interference management, has always been attractive to both academia and industry, as exemplified by coordinated multi-point (CoMP) \cite{36.219}, multiple transmission and reception points (mTRPs) \cite{38.214}, and cell-free multiple-input multiple-output (MIMO) \cite{ngo2017cell} networks.
Regarding sixth-generation (6G) networks, it is envisioned to further enhance the flexibility of multi-BS cooperation to satisfy the varying spatiotemporal traffic demands and meet diverse user experience requirements \cite{shen2017flexible, sun2023flexible}. 
Unlike conventional cooperation which typically employs fixed cooperation sets, flexible cooperation utilizes all potential BSs with arbitrary transmission modes (TMs) for achieving specific objectives such as spectral efficiency (SE) maximization and energy saving.

However, implementing flexible cooperation confronts significant challenges since the architecture of the fifth-generation (5G) radio access network (RAN) still follows the original single-cell based design \cite{chen2024evolution}.
First, the coupling between uplink (UL) and downlink (DL) transmissions restricts the flexibility of BS association and resource utilization.
For example, in time division duplex (TDD) systems, cooperating BSs are required to maintain consistent UL and DL frame structures, and it is not feasible to independently configure different cooperation sets for UL and DL transmissions.
Moreover, the performance gains of cooperation heavily depend on the accuracy and timeliness of channel state information (CSI), which can only be ensured in systems with tightly coupled UL and DL.
Second, the overheads of pilots and CSI feedback scale linearly with the number of cooperating BSs and users, reducing the effective data rate, as flexible cooperation requires CSI for an arbitrary pair of BS and user.
Furthermore, the computational complexity of determining joint transmission parameters for multi-BS cooperation increases exponentially  \cite{lee2014combinatorial}.
As a result, practical cooperation is generally constrained to two fixed BSs \cite{irmer2011coordinated}.
Third, inherent feedback delays in transmission lead to outdated CSI in the practical scheduling process, resulting in a mismatch between the cooperative resource scheduling decision and its desired objectives, e.g., meeting specific users' demands.

To enable flexible resource cooperation for 6G, we consider the fully-decoupled RAN (FD-RAN) architecture \cite{wang2023road}. 
It decouples BSs into control BSs (C-BSs) and data BSs, with data BSs further split into downlink BSs (DL-BSs) and uplink BSs (UL-BSs) \cite{yu2019fully}. 
By fully separating control, UL, and DL operations, FD-RAN offers enhanced flexibility to meet the diverse requirements of 6G \cite{qian2023enabling, liu2023leveraging, xu2023federated}.
At the physical layer, the feedback-free MIMO transmission scheme is employed, where transmission parameters are determined solely based on the user's geolocation, leveraging a mapping learned from the historical channel data.
Previous studies have demonstrated the feasibility and evaluated the performance of such kinds of methods \cite{liu2025enabling, xu2024fully}.
As for resource scheduling, DL and UL are also conducted independently, and users' location as well as the corresponding transmission parameters are used for obtaining scheduling decisions, thus eliminating the reliance on CSI such that flexible cooperation among arbitrary BSs can be potentially realized.

However, there still lacks existing research on feedback-free resource scheduling towards flexible multi-BS cooperation.
To this end, we investigate the problem of DL cooperation set selection and subcarrier allocation in FD-RAN.
The flexibility of resource allocation lies in that users can be associated with any potential DL-BSs operating in appropriate TMs. 
However, such flexibility also significantly increases the complexity of the problem, posing challenges to achieving optimal solutions.
Furthermore, resource allocation can have diverse and even conflicting objectives, including network throughput, fairness, and energy efficiency (EE).
Thus, appropriate scheduling objectives are required for various conditions to better utilize resources. 
Moreover, to align with the practical scheduling process, resource scheduling should meet stringent running time requirements, e.g., at the ms level. 
Consequently, designing resource allocation algorithms that achieve both high computational efficiency and scheduling quality in FD-RAN remains a very challenging problem.

To address the above challenges, we propose a feedback-free cooperative resource scheduling approach that can leverage the flexibility to achieve tailored scheduling objectives based on the current network load condition.
We also consider resource scheduling from the user’s perspective, focusing on meeting personalized service demands.
Specifically, the network conditions are classified into heavy-load and light-load based on whether all user demands can be met through JT of all DL-BSs, which provides the highest data rate per subcarrier. 
In heavy-load conditions, where user demands cannot be fully met, JT is still employed to maximize demand satisfaction. 
A greedy subcarrier allocation algorithm considering user service priorities and fairness is proposed and proves optimal.
In light-load conditions, where at least one solution exists to satisfy all user demands by JT of all DL-BSs, the focus shifts to finding the most energy-efficient cooperation strategy.
An efficient two-stage resource allocation (TSRA) algorithm is proposed, in which the branch-and-cut method \cite{mitchell2002branch} is employed to obtain the optimal solution. To improve computational efficiency, the feasibility pump algorithm \cite{fischetti2005feasibility} is incorporated to quickly generate a high-quality feasible solution as initialization.
To conclude, the contributions of this paper are listed as follows:
\begin{itemize}
    \item 
    To the best of our knowledge, this is the first study to investigate the feedback-free DL cooperative resource scheduling, fully leveraging the flexibility of FD-RAN.
    User location information instead of CSI is required for resource scheduling, eliminating the overheads of pilots and CSI feedback while providing insensitivity to transmission delays and scalability to the number of cooperating DL-BSs.
    \item 
    We formulate the problem of joint cooperation set selection and subcarrier allocation based on network load conditions.
    The scheduling is user-centric in that user demands are satisfied with the best effort.
    In heavy-load conditions, resource allocation aims to maximize the overall user demand satisfaction rate while considering different user priorities and maintaining fairness.
    In light-load conditions, the focus shifts to minimizing energy costs while still meeting user demands.
    \item 
    To solve the problems, the resource scheduler is designed with two tailored algorithms.
    In heavy-load conditions, a greedy subcarrier allocation algorithm is proposed and proves optimal.
    In light-load conditions, the TSRA algorithm is proposed to enable flexible and energy-efficient cooperation by providing a near-optimal solution within a short time.
    \item 
    We conduct extensive simulations on ray-tracing channel data.
    Simulation results validate the superiority of the proposed algorithms in terms of both performance and running time.
    Comparisons with feedback-based resource scheduling highlight the potential of the feedback-free approach for enabling flexible cooperation in practical networks. 
\end{itemize}

The remainder of this paper is organized as follows.
Related works are introduced in Section \ref{RW}.
The system model and problem formulation are given in Section \ref{SM and PF}.
Resource scheduling algorithms under heavy-load and light-load network conditions are presented in Section \ref{Heavy-Load} and Section \ref{Light-Load}, respectively.
Simulation results and discussions are given in Section \ref{SR and D}.
Finally, Section \ref{Con} concludes the whole paper.

\section{Related Works}
\label{RW}
\subsection{Feedback-free Transmission}
Due to the issues caused by CSI feedback, numerous studies have investigated methods to eliminate it \cite{vasisht2016eliminating, chikha2022radio, wang2023deep, jiwei2024channel, zeng2021toward, liu2021fire}.
In \cite{vasisht2016eliminating}, the R2-F2 system is introduced, enabling BSs to predict the DL channel based on the estimated UL channel. 
R2-F2 establishes a channel-to-path transform that captures the physical propagation paths, allowing inference of channels across different frequency bands and achieving a beamforming gain of 0.7 dB.
The radio environment map (REM) for a BS is introduced in \cite{chikha2022radio}, which is constructed using measurements across all grids of beams. 
The derived reference signal received powers from various beamforming schemes based on these REMs facilitate the interference coordination between the serving and neighboring cells.
The authors in \cite{wang2023deep} present a deep learning-based radio map that leverages user location information to determine beamforming vectors. 
The neural network loss function can be designed to accommodate diverse communication requirements.
In \cite{jiwei2024channel}, a radio map-based complex-valued precoding network (RMCPNet) is proposed to enable feedback-free transmission in FD-RAN. 
This method leverages location information along with statistical channel data available at the edge cloud, including transmit and receive correlation matrices as well as channel energy matrices, to improve transmission performance.
The concept of channel knowledge map (CKM) is first introduced in \cite{zeng2021toward} to support environment-aware communications. 
CKM employs the locations of transmitters and receivers to provide extra channel knowledge for transmission, with practical applications like beam selection, where the channel path map (CPM) reconstructs the channel based on the three most dominant paths.
In \cite{liu2021fire}, the FIRE system is proposed as an end-to-end approach using a variational autoencoder to acquire DL channel in frequency division duplex (FDD) systems without relying on user feedback. 
Instead of reconstructing input data, FIRE maps UL channels to DL channels, allowing BSs to directly obtain DL information.
Different from the above works, the considered feedback-free transmission scheme only requires user locations to determine MIMO multi-stream transmission parameters. 
Moreover, we address the resource scheduling problem to fully unlock the flexibility of FD-RAN.

\subsection{Resource Allocation Towards Multi-BS Cooperation}

Over the past decade, substantial efforts have been dedicated to resource allocation aimed at enabling multi-BS cooperation across diverse network settings and objectives\cite{yu2013multicell, zheng2014optimal, wang2015energy, he2019joint, qiao2023joint, zhang2024multi}.
\cite{yu2013multicell} proposes a coordinated scheduling, beamforming, and power allocation scheme across multiple BSs in wireless cellular networks. 
The study primarily aims to mitigate inter-cell interference and enhance overall network utility through the joint optimization of transmission strategies and resource allocation, utilizing uplink-downlink duality, interference pricing, and a heuristic joint proportional fair scheduling algorithm.
In \cite{zheng2014optimal}, a novel BS coordination approach to mitigate inter-cell interference in orthogonal frequency-division multiple access (OFDMA) based cellular networks is proposed. 
By formulating the resource allocation problem as a local cooperation game among BSs, the authors demonstrate the existence of a joint-strategy Nash equilibrium that leads to global optimality of network utility. 
In \cite{wang2015energy}, the authors investigate an energy-efficient resource allocation algorithm for multicell OFDMA systems with imperfect CSI.
The proposed algorithm aims to maximize the system's EE by jointly optimizing user scheduling, data rate adaptation, and power allocation.
To address co-channel interference, multiple BSs coordinate their resource allocation strategies based on shared CSI.
\cite{he2019joint} proposes a deep reinforcement learning (RL) framework for joint power allocation and channel assignment in non-orthogonal multiple access (NOMA) systems, where two users share the same channel resources. 
To optimize resource allocation, the authors develop RL with an attention-based neural network (ANN) to perform channel assignment and derive a closed-form solution for power allocation. 
\cite{qiao2023joint} investigates the joint optimization of resource allocation and user association in a heterogeneous cellular network (HCN) assisted by reconfigurable intelligent surface (RIS). 
The study focuses on a multi-BS multi-frequency network scenario, including 4G, 5G, millimeter wave (mmWave), and terahertz networks, with the goal of maximizing the system sum rate. 
To address the NP-hard and coupling relationship between RIS phase shift and resource allocation, the authors propose a block coordinate descent (BCD) method to alternately optimize local solutions and obtain the global solution. 
In \cite{zhang2024multi}, the maximization of SE in the cooperative low Earth orbit (LEO) multi-satellite network is studied, emphasizing the joint optimization of hybrid beamforming and user scheduling. 
In scenarios involving multiple ground users (GUs) and satellites, the proposed approach enhances SE by optimizing satellite-to-GU associations and mitigating interference using digital beamforming techniques.
Different from the above works, we investigate feedback-free DL resource scheduling to facilitate multi-BS flexible cooperation in FD-RAN.
Additionally, the scheduling process prioritizes meeting users' personalized demands and flexibly adapts to different objectives—incorporating priority and fairness under heavy-load conditions, and minimizing energy costs under light-load conditions.

\section{System Model and Problem Formulation}
\label{SM and PF}
In this section, we first introduce the system model of DL feedback-free MIMO transmission and resource scheduling in FD-RAN.
Then, we analyze and formulate the problem of feedback-free resource scheduling.

\subsection{System Model}

\begin{figure}[t]
    \begin{center}
        \includegraphics[width=0.43\textwidth]{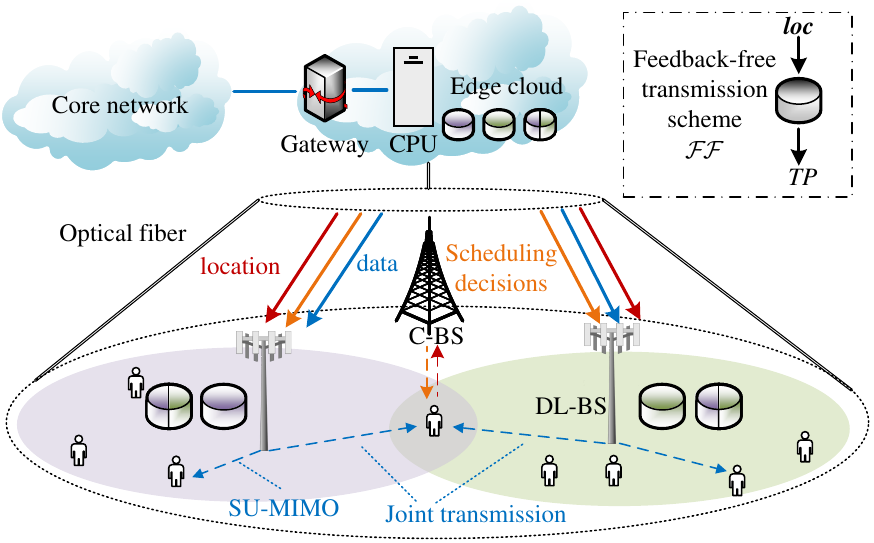}
        \caption{The multi-BS cooperation in DL network of FD-RAN.}
        \label{scneario}
    \end{center}
\end{figure} \par

We consider the DL network of FD-RAN illustrated in Fig. \ref{scneario}, where $M$ DL-BSs serve $N$ users.
In FD-RAN, feedback-free MIMO transmission is employed, where transmission parameters including the precoder, the number of spatial streams, and channel quality indicator (CQI) are determined based on the user's geolocation through a mapping derived from historical channel data.
This mapping, denoted as $ \mathcal{FF}_{\text{TM}}: \bm{loc} \rightarrow TP $, relates the user's 3-dimensional Cartesian coordinates $\bm{loc}$ to transmission parameters $TP$ according to the employed TM.
TMs, such as transmit diversity, single-user MIMO (SU-MIMO), and JT, have defined various transmission methods to optimize network performance under different scenarios.
In our previous work \cite{liu2025enabling}, we have studied how to establish such mapping, while other studies have investigated alternative approaches \cite{xu2024fully, liu2024leveraging}.
The DL network of FD-RAN operates over a specific DL band containing $K$ orthogonal subcarriers, each of which can be used by any DL-BS to serve users. 
Users demanding data report their locations to the C-BS. 
The edge cloud collects user locations from the C-BS, makes scheduling decisions, and sends them to DL-BSs and C-BS over high-capacity, low-latency optical fibers.
Then, the C-BS informs users of their assigned subcarriers via DL control information (DCI) through dedicated wireless bands, enabling the DL-BSs to transmit and the users to receive data on the specified subcarriers.
DL-BSs are also informed of users' locations, based on which they determine the appropriate precoders as well as modulation and coding schemes (MCSs) from $\mathcal{FF}_{\text{TM}}$.

\begin{figure}[t]
    \begin{center}
        \includegraphics[width=0.485\textwidth]{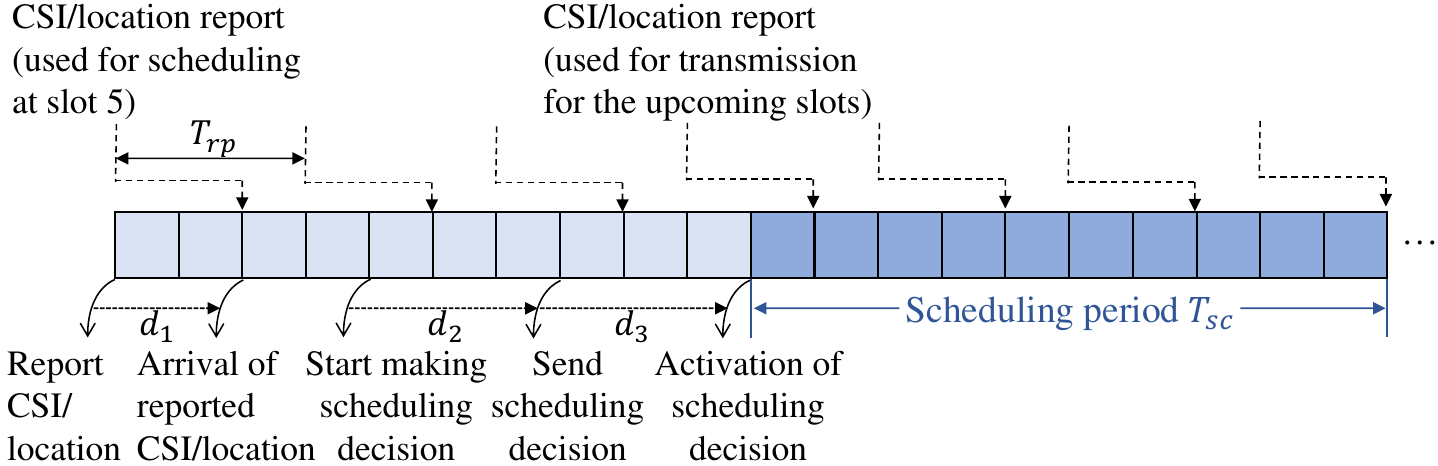}
        \caption{Illustration of the resource scheduling process in DL network of FD-RAN or 5G FDD systems.}
        \label{scheduling_process}
    \end{center}
\end{figure} \par

Fig. \ref{scheduling_process} presents the resource scheduling process over multiple time slots in the DL network of FD-RAN or 5G FDD systems.
Different from FD-RAN, the process in 5G starts with the user estimating the channel and reporting CSI.
In this paper, we take into account the delays associated with the reporting of location/CSI, the running time of the scheduling algorithm, and the activation of scheduling decisions, denoted as $d_1$, $d_2$, and $d_3$, respectively.
As defined by $k_0$, which specifies the delay between DCI and the DL data transmission slot in the 3GPP standard \cite{38.214}, $d_3$ must be no less than the maximum delay for control signaling to reach all users such that scheduling decisions can take effect simultaneously.
The location/CSI reporting period and scheduling period are denoted as $T_{rp}$ and $T_{sc}$, respectively.
Data transmissions depend on the previously activated scheduling decisions, while new decisions are made based on the latest received locations/CSI from the current time slot.
For 5G, it can be observed that due to various delays, CSI used in actual transmission may significantly differ from that used in the most recent scheduling decision, potentially affecting the scheduling performance.
Although location reporting in FD-RAN is also subject to delays, user locations generally remain stable within the ms-level scheduling period. 
Furthermore, FD-RAN employs fixed transmission parameters for each location, making it insensitive to transmission delays.

In this paper, we consider two TMs at the physical layer: SU-MIMO and multi-BS JT on the same subcarrier.

\subsubsection{SU-MIMO}
In SU-MIMO, a DL-BS equipped with $N_{tx}$ antennas transmits signals to a user with $N_{rx}$ antennas.
The received signal on the assigned subcarrier $k$ is given by
\begin{equation}
    \bm{c}_{k} = \mathbf{H}_{k}\mathbf{W} \bm{s}_{k} + \bm{u}_{k}, \quad k \in \mathcal{K}',
\end{equation}
where $\mathbf{H}_{k} \in \mathbb{C}^{N_{rx} \times N_{tx}}$ is the channel matrix that the transmitted symbol vector $\bm{s}_{k} \in \mathbb{C}^{L \times 1}$ with normalized unit power experiences on the subcarrier $k$.
$\mathbf{W} \in \mathbb{C}^{N_{tx} \times L}$ is the employed precoding matrix derived from $\mathcal{FF}_{\text{SU-MIMO}}$ and remains the same across all subcarriers.
$L$ denotes the number of layers employed for spatial multiplexing.
$\bm{u}_{k} \sim \mathcal{CN}(\bm{0}, \sigma^2)$ is the additive white Gaussian noise (AWGN) with variance $\sigma^2$.
$\mathcal{K}' \subseteq \mathcal{K} = \{1, \ldots, K\}$ denotes the set of allocated subcarriers from all available subcarriers $\mathcal{K}$.
Then SINR of $\ell$-th layer after utilizing equalizer $\mathbf{E}_{k} \in \mathbb{C}^{L \times N_{rx}}$ on $ \mathbf{H}_{k}\mathbf{W}$ is given as
\begin{equation}
    \label{SINR}
    \text{SINR}_{k,\ell} = \frac{\|(\mathbf{G}_{k})_{\ell,\ell}\|_{F}^2}
    {\sum_{i=1, i \neq \ell}^{L} \|(\mathbf{G}_{k})_{\ell,i}\|_{F}^2 + \sigma^2_n \sum_{i=1}^{N_{rx}} \|(\mathbf{E}_{k})_{\ell,i}\|_{F}^2},
\end{equation}
where $k \in \mathcal{K}', \ell \in \mathcal{L} = \{1,\ldots,L\}$ and $\|\cdot\|_{F}$ denotes the Frobenius norm.
$\mathbf{G}_{k} = \mathbf{E}_{k}\mathbf{H}_{k}\mathbf{W}$ is the equivalent channel matrix, and $(\cdot)_{\ell,i}$ denotes the element in the $\ell$-th row and $i$-th column.
Thus, the numerator of (\ref{SINR}) represents the desired signal power of layer $\ell$ on subcarrier $k$.
The first term of the denominator refers to the inter-layer interference and the second term is the enhanced noise.

The post-equalization SINRs across all assigned subcarriers $\mathcal{K}'$ and all layers $\mathcal{L}$ are transformed into an effective SNR of an equivalent single-input single-output (SISO) system with an AWGN channel \cite{brueninghaus2005link}. 
The transformation is defined as
\begin{equation}
    \label{effSNR}
    \text{SNR}_{\text{eff}} = \alpha f^{-1}\left(\frac{1}{|\mathcal{K}'|L} \sum_{k \in \mathcal{K}'} \sum_{\ell \in \mathcal{L}} f\left(\frac{\text{SINR}_{k,\ell}}{\alpha}\right)\right),
\end{equation}
where $f$ denotes the transformation function and $f^{-1}$ represents its inverse. 
The mutual information effective SNR mapping (MIESM) is employed, where $f$ corresponds to the bit-interleaved coded modulation (BICM) capacity \cite{caire1996capacity}. 
The parameter $\alpha$ serves as an adjustment factor, tuning the block error rate (BLER) performance of the SISO channel to closely approximate that of the MIMO channel.
Based on $\text{SNR}_{\text{eff}}$, the real CQI value $\text{CQI}_{re}$ corresponding to the employed precoder is selected as the highest MCS that maintains the BLER below 0.1. 
A high MCS value indicates the use of high-order modulation and a high coding rate scheme, resulting in increased SE and throughput.

Note that due to the time-varying channels, the employed CQI from $\mathcal{FF}_{\text{SU-MIMO}}$ $\text{CQI}_{em}$ may vary from the real CQI value.
Thus, the employed modulation order $\Theta$ and coding rate $R_{c}$ associated with $\text{CQI}_{em}$, which can be obtained from the CQI table in \cite{36.213}, should be considered into the calculation of the user's achievable data rate.
Furthermore, a higher employed CQI than real CQI can result in a BLER of up to 100\%, leading to zero throughput. 
Therefore, the user's achievable data rate in bit per second can be expressed as follows, in accordance with the 3GPP standard \cite{38.306}:
\begin{equation}
    \label{data rate}
    R = 
    \begin{cases}
     L \Theta R_{c} \frac{|\mathcal{K}'|}{t^{\mu}} (1-OH),&{\text{if}}\ \text{CQI}_{em} \leqslant \text{CQI}_{re},\\
     {0,}&{\text{if}}\ \text{CQI}_{em} > \text{CQI}_{re},
    \end{cases}
\end{equation}
where $t^{\mu} = \frac{10^{-3}}{14 \times 2^{\mu}}$ is the average orthogonal frequency-division multiplexing (OFDM) symbol duration in a time slot for numerology $\mu$.
The subcarrier spacing in kHz is given by $2^{\mu} \times 15$.
$|\mathcal{K}'|$ denotes the number of allocated subcarriers.
$OH$ represents the overhead ratio for control signaling, synchronization, and demodulation reference signals.
For simplicity, we assume that selecting a higher CQI results in zero throughput. 
For more realistic data rates, the high-fidelity Vienna 5G link-level simulator can generate a table of BLERs for different CQI choices \cite{pratschner2018versatile}, enabling the calculation of actual rates as original rates multiplied by $(1-\text{BLER})$.

\subsubsection{JT}
In JT of CoMP, multiple BSs simultaneously transmit the same signal to a user on the same subcarrier, enhancing the user's received signal strength.
The received signal on the assigned subcarrier $k$ from the cooperation set $\mathcal{M}' \subseteq \mathcal{M} = \{1, \ldots, M\}$ can be expressed as
\begin{equation}
    \bm{c}_{k} = \sum_{m \in \mathcal{M}'} \mathbf{H}_{k}^{m} \mathbf{W}^{m} \bm{s}_{k} + \bm{u}_{k}, \quad k \in \mathcal{K}',
\end{equation}
where $\mathbf{H}_{k}^{m}$ represents the channel from BS $m$ in the cooperation set and $\mathcal{M}$ denotes the available BS set.
The precoder $\mathbf{W}^{m}$ is derived from $\mathcal{FF}_{\text{JT}_{\mathcal{M}'}}$ of the JT performed by the cooperating BSs in $\mathcal{M}'$.
The equalizer $\mathbf{E}_{k} \in \mathbb{C}^{L \times N_{rx}}$ is then computed based on the combined channel $ \sum_{m \in \mathcal{M}'} \mathbf{H}_{k}^{m} \mathbf{W}^{m}$. 
Similarly, the true CQI is determined using the $\text{SNR}_{\text{eff}}$ obtained from the transformation of all SINRs.

\subsection{Problem Formulation}
In this paper, we study the joint cooperation set selection and subcarrier allocation problem.
We assume that the aforementioned feedback-free transmission schemes $\mathcal{FF}_{\mathcal{TM}}$ are established and stored at the edge cloud and DL-BSs, where $\mathcal{TM}$ represents the TM set of SU-MIMO and JT provided by all possible combinations of $M$ DL-BSs. 
Thus, $|\mathcal{TM}| = 2^{M}-1$.
Since subcarrier transmit power potentially impacts the number of spatial streams and CQI, a fixed power level is applied to each subcarrier for simplicity. 
Furthermore, we assume that each subcarrier is either exclusively allocated to a single user so as to avoid interference or left unused for energy-saving purposes.

Specifically, we define $x_{m,n}$ as the decision variable representing the number of subcarriers allocated by BS $m$ to user $n$, where $x_{m,n} \in \{0, 1, \ldots, K \}, m \in \mathcal{M}, n \in \mathcal{N} = \{1, \ldots, N\}$.
$x_{m,n}=0$ indicates that BS $m$ does not serve user $n$.
Let $\mathcal{M}_{n}$ denote the cooperation set of user $n$, where $\mathcal{M}_{n} = \{m | x_{m,n}>0, \forall m \in \mathcal{M}\}$.
If $|\mathcal{M}_{n}|=1$, it represents the SU-MIMO TM.
In the case of multi-BS cooperation, i.e., when $|\mathcal{M}_{n}|>1$, if for any $m, m' \in \mathcal{M}_{n}$ it holds that $x_{m,n}=x_{m',n}$, then all BSs in the cooperation set allocate the same number of subcarriers to serve user $n$, indicating the adoption of JT.
Otherwise, a hybrid TM combining SU-MIMO and JT is employed.

Depending on the number of subcarriers allocated by multiple BSs in the cooperation set, the hybrid TM can encompass various cases, ranging from a single BS to two cooperating BSs serving the user, and extending to all BSs in the cooperation set jointly serving the user.
For instance, suppose the cooperation set is $\{1, 2, 3\}$, and the number of subcarriers allocated by the BSs in the cooperation set is $\{8, 2, 4\}$.
This subcarrier allocation scheme will include the following TMs:
First, BSs 1, 2, and 3 jointly use 2 subcarriers for three-BS JT;
Then, BSs 1 and 3 use additional $4-2=2$ subcarriers to perform two-BS JT;
Finally, BS 1 uses the remaining $8-4=4$ subcarriers to serve the user independently.
If the number of subcarriers allocated by the BSs in the cooperation set is $\{8, 4, 4\}$, then only the three-BS JT on 4 subcarriers and BS 1 serving the user independently with the remaining $8-4=4$ subcarriers are included in the TMs.
With all decision variables determined, resource mapping is performed by allocating subcarriers to each user in turn from the first subcarrier in $\mathcal{K}$, in accordance with the aforementioned procedure.

The primary objective of resource scheduling is to meet the diverse data rate demands of all users as much as possible.
Since the JT mode can improve data rates without requiring additional subcarriers by enabling multiple BSs to serve users using the same frequency resources, we first attempt to achieve this objective through the JT of all DL-BSs. 
If it remains impossible to satisfy the demands of all users, this indicates that the network is in a heavy-load condition, and spectrum resources are in shortage.
In this case, the objective of resource scheduling should turn to consider users' priorities while maintaining fairness by meeting a minimum level of demand for all users.

Specifically, we define $R_{n}^{d}$ as the data rate demand of user $n$ and $\hat{R}_{n}$ as the expected achievable rate after resource allocation.
The demand satisfaction rate of user $n$ can then be expressed as $\eta_n = \min \{ \frac{\hat{R}_{n}}{R_{n}^{d}}, 1 \}$.
Clearly, the maximum demand satisfaction rate is 100\%, and allocating subcarriers beyond the required demand leads to waste in frequency and power resources without additional benefit.
The priority weight of user $n$, denoted as $\omega_n$, reflects the priority of the user's service.
We consider normalized priority weights such that $\sum_{n \in \mathcal{N}} \omega_n = 1$.
In addition, we define a minimum demand satisfaction rate $\eta_{\min}$ as a fairness constraint to prevent low-priority users from being entirely excluded from resource scheduling.
Thus, under heavy-load network conditions, the studied resource scheduling problem can be formulated as
\begin{alignat}{2}
\max_{x_{m,n}} \quad & \sum_{n \in \mathcal{N}} v_n = \sum_{n \in \mathcal{N}} \omega_n \eta_n \label{objective1} &\\
\mbox{s.t.}\quad
&x_{m,n} \in \{0, 1,\ldots, K \},  \forall m \in \mathcal{M}, \forall n \in \mathcal{N}, &\\
& \sum_{n \in \mathcal{N}}\max_{m \in \mathcal{M}}\{x_{m,n}\} \leqslant K, \label{subcarrier}&\\
& x_{m,n} = x_{m',n},\quad  \forall m, m' \in \mathcal{M}, \forall n \in \mathcal{N}, \label{CoMP}&\\
& \eta_{\min} \leqslant \eta_n, \quad \forall n \in \mathcal{N}, \label{lower limit}&\\
& \mathcal{FF}_{\text{JT}_{\mathcal{M}}}:\bm{loc}_{n} \rightarrow TP_{n}, \quad \forall n \in \mathcal{N}. \label{feedback-free} &
\end{alignat}
$v_{n}$ denotes the weighted demand satisfaction rate of user $n$.
Constraint (\ref{subcarrier}) ensures that the total number of allocated subcarriers remains within the subcarrier number limit.
TM is restricted to the JT of all DL-BSs indicated by (\ref{CoMP}) and (\ref{feedback-free}), as it is the most effective way to meet user demands in heavy-load network conditions.
% Constraint (\ref{lower limit}) defines the minimum acceptable user demand satisfaction rate.
% Finally, (\ref{feedback-free}) indicates the feedback-free transmission modes established at the physical layer, mapping location $\bm{loc}$ to transmission parameters $TP$.

If the demands of all users can be satisfied through the JT of all DL-BSs, then spectrum resources are relatively abundant.
In this case, the primary objective of resource scheduling has been achieved. 
Next, we should focus on minimizing the subcarrier energy cost while meeting all user demands, thereby reducing operational expenditures for operators.
Assuming the transmit power per subcarrier is $p$, the total subcarrier energy cost can be expressed as $e = p \sum_{m \in \mathcal{M}} \sum_{n \in \mathcal{N}} x_{m,n}$.
Thus, the studied resource scheduling problem in light-load network conditions can be formulated as follows:
\begin{alignat}{2}
\min_{x_{m,n}} \quad & e = p \sum_{m \in \mathcal{M}} \sum_{n \in \mathcal{N}} x_{m,n}  \label{objective2}&\\
\mbox{s.t.}\quad
& x_{m,n} \in \{0, 1,\ldots, K \}, \forall m \in \mathcal{M}, \forall n \in \mathcal{N}, &\\
& \sum_{n \in \mathcal{N}}\max_{m \in \mathcal{M}}\{x_{m,n}\} \leqslant K, &\\
& R_{n}^{d} \leqslant \hat{R}_{n} ,\quad \forall n \in \mathcal{N}, \label{demand} &\\
& \mathcal{FF}_{\mathcal{TM}}:\bm{loc}_{n} \rightarrow TP_{n}, \quad \forall n \in \mathcal{N}.  &
\end{alignat}
Constraint (\ref{demand}) ensures that data rate demands of all users are met.
To minimize the subcarrier energy cost while meeting user demands, all considered TMs are feasible options for each user in light-load conditions. 
Therefore, no specific constraints on the TM, such as (\ref{CoMP}) and (\ref{feedback-free}), are applied in this case.

\section{Resource Scheduling Under Heavy-Load Condition}
\label{Heavy-Load}
Under heavy-load network conditions, the TM is fixed to JT by all DL-BSs, i.e., all DL-BSs allocate the same number of subcarriers to a given user. 
Thus, the decision variable for user $n$ can be simplified as $x_{n}, \forall n \in \mathcal{N}$. 
This leads to a significant reduction in the solution space, from $(MN)^{K}$ to $N^{K}$, making the problem more tractable.
To maximize the network's weighted sum of user demand satisfaction rates while meeting the minimum demand satisfaction rate constraint, we propose a greedy algorithm to solve the problem.

After allocating the minimum number of subcarriers to meet user demand threshold $\eta_{\min}$, we can collect a set of user information, including the user ID, priority weight $\omega$, the expected achievable data rate $\hat{R}^{1}$ after allocating one subcarrier, the data rate demand $R^{d}$, and the required number of subcarriers $K^{d}$.
The total number of subcarriers remaining for scheduling is denoted by $K_{\text{remaining}}$.
$\hat{R}^{1}$, which depends on the user location, can be obtained from the established $\mathcal{FF}_{\text{JT}_{\mathcal{M}}}$.
As subcarriers can only be allocated in integer values, the required number of subcarriers for user $n$ is determined by rounding up, i.e., $K^{d} = \lceil \frac{R^{d}}{\hat{R}^{1}} \rceil$.
The goal is to maximize the weighted sum of user demand satisfaction rates while staying within the available subcarrier number limit.
Thus, subcarriers should be allocated to users with the highest satisfaction rate gain per additional subcarrier.
Here, we introduce a greedy weight $\vartheta_{n}$ as the ranking criterion, which represents the expected weighted demand satisfaction rate when scheduling one subcarrier to user $n$: 
\begin{equation} 
\label{greedy weight} 
\vartheta_{n} = 
\begin{cases} \omega_{n} \frac{\hat{R}_{n}^{1}}{R_{n}^{d}}, &K_{n}^{d} > 1,\\ 
\omega_{n}(1-\frac{(K_{n}^{d}-1) \times \hat{R}_{n}^{1}}{R_{n}^{d}}), &K_{n}^{d} = 1. 
\end{cases} 
\end{equation} 
It is worth noting that if $K_{n}^{d} = 1$, the weighted demand satisfaction rate from allocating one additional subcarrier will be lower than that in cases where the subcarrier demand is greater than 1. 
This is because the maximum satisfaction rate for a user is 100\%, and in this case, the allocated subcarrier would cause the data rate to exceed the user's demand.

Given that the greedy weight of a user changes with the number of allocated subcarriers, we divide the user $n$'s information into two parts: one for the demand of the first $K_{n}^{d} - 1$ subcarriers and the other for the demand of the last subcarrier. 
Therefore, the problem $\mathcal{Z}$ under heavy-load network conditions can be expressed as 
\begin{equation}
    \label{greedy problem}
    \begin{split}
        \mathcal{Z}  = & \bigcup_{\forall n \in \mathcal{N}} \{\mathcal{Z}_{n}\}
         = \bigcup_{\forall n \in \mathcal{N}} \{\mathcal{Z}_{n}^{:K_{n}^{d}-1}\} \cup \{\mathcal{Z}_{n}^{-1}\}\\ 
        % = &\bigcup_{\forall n \in \mathcal{N}}\{(n,\vartheta_{n}, K_{n}^{d}-1 ) \} \cup \{(n,\vartheta_{n}, 1 )\}  \\
        = &\bigcup_{\forall n \in \mathcal{N}}\{(n,\omega_{n} \frac{\hat{R}_{n}^{1}}{R_{n}^{d}}, K_{n}^{d}-1 ) \} \cup \\ 
        & \{ (n,\omega_{n}(1-\frac{(K_{n}^{d}-1) \times \hat{R}_{n}^{1}}{R_{n}^{d}}), 1 )\}.
    \end{split}
\end{equation}
$\mathcal{Z}_{n}$ denotes the subproblem of user $n$, consisting of the user ID, greedy weight, and subcarrier demand.
$\mathcal{Z}_{n}^{:K_{n}^{d}-1}$ refers to the subproblem for the first $K_{n}^{d}-1$ subcarriers, and $\mathcal{Z}_{n}^{-1}$ corresponds to the last subcarrier demand.
The subcarrier allocation solution for $\mathcal{Z}$ is defined as $\mathcal{X} = \bigcup_{n \in \mathcal{N}}\{x_{n}\} = \bigcup_{n \in \mathcal{N}}\{x_{n}^{:K_{n}^{d}-1} + x_{n}^{-1}\}$, where $x_{n}^{:K_{n}^{d}-1}$ and $x_{n}^{-1}$ represent the allocations for the first $K_{n}^{d}-1$ subcarriers and the last subcarrier, respectively.
Similarly, the weighted sum of user demand satisfaction rates is $V = \sum_{n \in \mathcal{N}} v_{n} = \sum_{n \in \mathcal{N}} ( v_{n}^{:K_{n}^{d}-1} + v_{n}^{-1} )$, where $v_{n}^{:K_{n}^{d}-1}$ and $v_{n}^{-1}$ denote the satisfaction rates from $x_{n}^{:K_{n}^{d}-1}$ and $x_{n}^{-1}$, respectively.

The greedy algorithm can be described as follows: at each step, we select the user with the highest greedy weight and allocate as many subcarriers as possible to that user. 
We repeat this process until all remaining subcarriers are allocated. 
The algorithm's complexity mainly arises from the aforementioned procedure and updating the user queue. 
The time complexity is $\mathcal{O}(N \log N)$ if a heap-based structure (e.g., priority queue) is employed to manage users' greedy weights.
Updating the heap after each allocation involves $\mathcal{O}(\log N)$, and in the worst case, all $N$ users are processed.
The space complexity is $\mathcal{O}(N)$, as the algorithm requires to store a sorted information table for all users.
In what follows, we demonstrate the optimality of the proposed greedy algorithm by proving the greedy choice property and showing the optimal substructure.

\begin{proof}
Let user $n$ be the one with the highest greedy weight. 
The goal is to prove that the optimal solution involves allocating as many subcarriers as possible to user $n$.
We prove that this statement is true by contradiction.
Suppose there exists an optimal solution in which we do not allocate as many subcarriers as possible to user $n$, and all remaining subcarriers have already been allocated. 
Since user $n$ has the highest greedy weight, there must exist another user $n'$ such that $\vartheta_{n'} < \vartheta_{n}$.
We can take $\Delta$ subcarriers from the allocation to user $n'$ and schedule them to user $n$. 
The change in the weighted sum of user demand satisfaction is given by $\Delta \vartheta_{n} - \Delta \vartheta_{n'} = \Delta (\vartheta_{n} - \vartheta_{n'}) > 0$, since $\vartheta_{n'} < \vartheta_{n}$.
This leads to a contradiction because the “optimal” solution assumed could be improved by reallocating some subcarriers from user $n'$ to user $n$. 
Hence, it is not optimal.
\end{proof}

\begin{proof}
Suppose that $\mathcal{X}$ is the optimal solution to the problem $\mathcal{Z}$ with $K_{\text{remaining}}$ subcarriers, and it has the optimal weighted sum of demand satisfaction rates $V$.
Our goal is to prove that $\mathcal{X}' = \mathcal{X} \setminus \{x_{n}^{:K_{n}^{d} - 1}\}$ is the optimal solution to the subproblem $\mathcal{Z}' = \mathcal{Z} \setminus \{\mathcal{Z}_{n}^{:K_{n}^{d} - 1}\}$, which has a weighted sum of demand satisfaction rates $V'$ and $K_{\text{remaining}}' = K_{\text{remaining}} - (K_{n}^{d} - 1)$ subcarriers available for allocation.
We prove this by contradiction. 
Suppose that $\mathcal{X}'$ is not the optimal solution to $\mathcal{Z}'$ and that there exists another solution $\mathcal{X}''$ which yields a higher weighted sum of demand satisfaction rates, i.e., $V'' > V'$.
Then, $\mathcal{X}'' \cup \{x_{n}^{:K_{n}^{d} - 1}\}$ is a feasible solution to the original problem $\mathcal{Z}$, and the corresponding weighted sum of demand satisfaction rates is $V'' + v_{n}^{:K_{n}^{d} - 1} > V' + v_{n}^{:K_{n}^{d} - 1} = V$.
This contradicts the assumption that $V$ is optimal.
\end{proof}

\section{Resource Scheduling Under Light-Load Condition}
\label{Light-Load}
Under light-load network conditions, the goal is to minimize the network's subcarrier energy cost while still meeting all user demands.
In this case, all considered TMs are available and the decision variable space is $(MN)^{K}$, making the direct solution to the problem computationally expensive.
To this end, we transform the original problem into a standard integer linear programming (ILP) problem and propose the two-stage resource allocation algorithm.

\subsection{Problem Transformation}
We modify the original decision variable to enable the transformation of the original optimization problem into an ILP problem.
Specifically, we define $\mathcal{M}^{q}$ as the collection of cooperative sets, each consisting of $q$ BSs selected from the BS set $\mathcal{M}$.
For example, when there are three BSs in the network, i.e., $\mathcal{M} = \{1, 2, 3\}$, the set of all combinations of two BSs selected from $\mathcal{M}$ is $\mathcal{M}^{2} = \{\{1, 2\}, \{1, 3\}, \{2, 3\}\}$.
Next, we introduce a binary variable $\delta_{\mathcal{B}, n}$ to represent the selection of the cooperation set $\mathcal{B}$ for user $n$. 
When $\delta_{\mathcal{B}, n} = 1$, it indicates that cooperation set $\mathcal{B} \in \mathcal{M}^{q}$ is selected to serve user $n$, where $q \in \mathcal{M}$ and $|\mathcal{B}| = q$; if $\delta_{\mathcal{B}, n} = 0$, it means the cooperation set is not chosen.
Let $o_{\mathcal{B}, n}$ denote the number of subcarriers allocated to user $n$ by cooperation set $\mathcal{B}$. 
The constraint that the total number of allocated subcarriers does not exceed the available total number of subcarriers can be transformed into
\begin{equation}
    \label{subcarrier_constraint}
    \sum_{n\in\mathcal{N}}\sum_{q \in \mathcal{M}}\sum_{\mathcal{B}\in \mathcal{M}^{q}} o_{\mathcal{B}, n} \leqslant K.
\end{equation}
In addition, we need to add the constraint 
\begin{equation}
    \label{selection_constraint}
    o_{\mathcal{B}, n} \leqslant \lambda \delta_{\mathcal{B}, n}
\end{equation}
to ensure that subcarriers are allocated only when the corresponding TM is selected. 
Here, $\lambda$ is a large constant that ensures this constraint is active only when $\delta_{\mathcal{B}, n} = 1$.

Based on the definition of the original decision variable, if a cooperation set of $q$ BSs is selected to serve a user, any cooperation involving fewer BSs must be chosen from the already selected $q$ BSs. 
Otherwise, allocating additional subcarriers could change the expected TMs.
For instance, if the cooperation set $\{1, 2\}$ has already scheduled one subcarrier to serve a user, then the single BS operating in the SU-MIMO mode must be selected from BS 1 or BS 2. 
If BS 3 is instead selected to allocate one additional subcarrier to the user, the expected TMs of two-BS JT combined with a single-BS service would transition to three-BS JT.
The TM compatibility constraint can be expressed as
\begin{equation}
    \label{compatibility_constraint}
    \delta_{\mathcal{B}, n} + \delta_{\mathcal{B}', n} \leqslant 1, \text{if } |\mathcal{B}'| < |\mathcal{B}| \text{ and } \mathcal{B}' \not\subset \mathcal{B}.
\end{equation}
Similarly, to avoid changing the TM, for each number of cooperating BSs $q$, only one BS combination can be selected to serve the user, or none at all:
\begin{equation}
    \label{combination_constraint}
    \sum_{\mathcal{B} \in \mathcal{M}^{q}} \delta_{\mathcal{B}, n} \leqslant 1, \forall q \in \mathcal{M}, \forall n \in \mathcal{N}.
\end{equation}

Based on the established $\mathcal{FF}_{\mathcal{TM}}$ at the physical layer, we can obtain the expected data rate per subcarrier provided by each cooperation set $\mathcal{B}$ for user $n$, denoted as 
$\hat{R}^{1}_{\mathcal{B}, n}$. 
The original optimization problem under light-load network conditions can then be transformed into
\begin{alignat}{2}
\min_{\delta_{\mathcal{B}, n}, o_{\mathcal{B}, n}} \quad & e = p \sum_{n\in\mathcal{N}}\sum_{q \in \mathcal{M}}\sum_{\mathcal{B}\in \mathcal{M}^{q}}  |\mathcal{B}| o_{\mathcal{B}, n}   \label{objective2-ILP}&\\
\mbox{s.t.}\quad
& \text{(\ref{subcarrier_constraint}), (\ref{selection_constraint}), (\ref{compatibility_constraint}), (\ref{combination_constraint})}, \notag &\\
& o_{\mathcal{B}, n} \in \{0, 1,\ldots, K \}, \forall \mathcal{B} \in \mathcal{M}^{q}, \notag &\\
& \qquad \qquad \qquad \qquad \quad \forall q \in \mathcal{M}, \forall n \in \mathcal{N}, &\\
& \delta_{\mathcal{B}, n} \in \{0, 1\}, \forall \mathcal{B} \in \mathcal{M}^{q}, \forall q \in \mathcal{M}, \forall n \in \mathcal{N}, &\\
& R_{n}^{d} \leqslant \sum_{q \in \mathcal{M}}\sum_{\mathcal{B}\in \mathcal{M}^{q}} o_{\mathcal{B}, n} \hat{R}^{1}_{\mathcal{B}, n}, \forall n \in \mathcal{N}. \label{new_demand} &
\end{alignat}
The objective function in (\ref{objective2-ILP}) is expressed in terms of the transformed integer decision variables, while constraint (\ref{new_demand}) ensures that user demands are satisfied.
% The objective function and all constraints are linear, with decision variables restricted to integers, resulting in a standard ILP problem.
The objective function and all constraints are linear, with decision variables constrained to integers, forming a standard ILP problem, denoted as $\mathcal{Z}_0$.

\subsection{Two-Stage Resource Allocation Algorithm}

To solve the aforementioned ILP problem, the TSRA algorithm is proposed. 
First, the feasibility pump algorithm quickly provides a high-quality feasible solution for subcarrier allocation. 
Then, the branch-and-cut method is applied to iteratively refine the solution to achieve optimality.

The feasibility pump algorithm is a heuristic method used to find feasible solutions for integer programming problems \cite{fischetti2005feasibility}. 
The goal is to obtain an integer feasible solution by iteratively constructing a new objective function based on the fractional solutions from the LP relaxation of the original ILP problem. 
The LP relaxation of the integer constraints on the decision variables can be expressed as
\begin{alignat}{2}
& 0 \leqslant o_{\mathcal{B}, n} \leqslant K, \forall \mathcal{B} \in \mathcal{M}^{q}, \forall q \in \mathcal{M}, \forall n \in \mathcal{N},  \label{relaxed_o}&\\
& 0 \leqslant \delta_{\mathcal{B}, n} \leqslant 1, \forall \mathcal{B} \in \mathcal{M}^{q}, \forall q \in \mathcal{M}, \forall n \in \mathcal{N}. \label{relaxed_delta}&
\end{alignat}
For simplicity, let index $j$ represent the combination of a possible cooperation set and user. 
Then, $o_{\mathcal{B}, n}$ and $\delta_{\mathcal{B}, n}$ can be rewritten as $o_{j}^{}$ and $\delta_{j}^{}$, where $j \in \mathcal{J} = \{1, \ldots, (2^{M}-1) \times N\}$.

LP problem has been proven to be solvable in polynomial time using the interior-point method \cite{karmarkar1984new}.
Denote the optimal solution to the LP problem as $\bm{\chi}^{\star} = \{\bm{o}^{\star}, \bm{\delta}^{\star}\}$.
If $\bm{\chi}^{\star}$ are already integers, a feasible integer solution to the original ILP problem $\mathcal{Z}_{0}$ is obtained. 
Otherwise, they are transformed into their integer values $[\bm{\chi}^{\star}] = \{[\bm{o}^{\star}], [\bm{\delta}^{\star}]\}$, where $[\cdot]$ denotes rounding to the nearest integer.
Note that $[\bm{\chi}^{\star}]$ may violate the original problem's constraints.
Thus, a new LP problem is formulated to adjust $\bm{\chi}$ to approximate the integer solution $[\bm{\chi}^{\star}]$ while meeting these constraints.
The distance between the linear and target integer solution is given by
\begin{equation}
\begin{aligned}
    & d(\bm{\chi}, [\bm{\chi}^{\star}])  = \sum_{[\delta_{j}^{\star}] =1 } (1-\delta_j) + \sum_{[\delta_{j}^{\star}] =0} \delta_j \\
    & + \sum_{[o_{j}^{\star}] =K} (K-o_j)
    + \sum_{[o_{j}^{\star}] =0} o_j
    + \sum_{0 < [o_{j}^{\star}] < K} (o_j^+ + o_j^-),
\end{aligned}
\end{equation}
where $o_j^+$ and $ o_j^-$ represent the non-negative differences between $o_j$ and $[o_j]$, corresponding to cases where $o_j > [o_j]$ and $o_j < [o_j]$, respectively.
Then, the formulated LP problem in feasibility pump $\mathcal{Z}_\text{LP}(\bm{\chi},[\bm{\chi}^{\star}])$ can be expressed as follows with additional constraints on $o_j$:
\begin{equation}
\label{FP_LP}
\begin{aligned}
    \min_{\bm{\chi}} \quad & d(\bm{\chi},[\bm{\chi}^{\star}])  \\
\mbox{s.t.}\quad & \text{(\ref{subcarrier_constraint}), (\ref{selection_constraint}), (\ref{compatibility_constraint}), (\ref{combination_constraint}), (\ref{new_demand}), (\ref{relaxed_o}), (\ref{relaxed_delta})}, \\
 & o_{j} = o_{j}^{\star} + o_j^+ - o_j^-, \quad \forall j \in \mathcal{J}: 0 < [o_{j}^{\star}] < K, \\
 & o_j^+, o_j^- \geqslant 0, \quad \forall j \in \mathcal{J}: 0 < [o_{j}^{\star}] < K.
\end{aligned}
\end{equation}
If $d(\bm{\chi},[\bm{\chi}^{\star}])  = 0$, a feasible solution to the original problem is obtained. 
The feasibility pump algorithm, forming the first stage of the proposed resource allocation method, is detailed in the initial part of Algorithm \ref{BC_FP}.
Note that in Step \ref{l10} of Algorithm \ref{BC_FP}, $N_{\text{flip}}$ elements of $\bm{\chi}^{*}$ are replaced with $\bm{\chi}^{\star}$ rounded in the opposite direction to prevent an endless loop.

\begin{algorithm}[t]
\label{BC_FP}
\SetKwInOut{Input}{Input}\SetKwInOut{Output}{Output}
\DontPrintSemicolon
\caption{TSRA.}
\LinesNumbered
\Input{Original ILP problem $\mathcal{Z}_{0}$}
\Output{Optimal solution $\bm{\chi}^{*}$ and optimal subcarrier energy cost $e^{*}$}
Initialize $\bm{\chi}^{*}$ as $\emptyset$ \label{l1}\; 
Solve the LP relaxation of the original problem to obtain $\bm{\chi}^{\star}$\; 

\While{$\bm{\chi}^{*}$ is not an integer solution}
{   
    \eIf{$\bm{\chi}^{\star}$ is an integer solution}
    {   
        $\bm{\chi}^{*} \leftarrow \bm{\chi}^{\star}$ \; 
    }
    {
        
        \eIf{$\exists j \in \mathcal{J}:  [\chi_j^{\star}] \neq \chi_j^{*}$}
        { 
            $\bm{\chi}^{*} \leftarrow [\bm{\chi}^{\star}]$ \; 
        }
        { 
            Select $N_{\text{flip}}$ variables of $\bm{\chi}^{*}$ with the highest $|\chi_j^{\star} - \chi_j^{*}|$ and replace $\chi_j^{*}$ with $\chi_j^{\star} $ rounded in the opposite direction \; \label{l10}
        }
        Solve $\mathcal{Z}_\text{LP}(\bm{\chi}^{\star}, \bm{\chi}^{*})$ of (\ref{FP_LP}) to obtain $\bm{\chi}^{\star}$ \;
    }
}
Initialize $e^{*}$ as $e(\bm{\chi}^{*})$ using the results from above\; \label{l15}
Initialize a set of subproblems $\mathcal{Y} \leftarrow \{\mathcal{Z}_{0}\}$ \;
\While{$\mathcal{Y} \neq \emptyset$ \label{l17}}
{
    Select and remove a subproblem $\mathcal{Z}_{y}$ from $\mathcal{Y}$\;
    Solve the LP relaxation of the problem $\mathcal{Z}_{y}$ to obtain $\bm{\chi}^{\star}$\ and subcarrier energy cost $e(\bm{\chi}^{\star})$ \; \label{l19}
    \eIf{$\bm{\chi}^{\star}$ is infeasible}
    {
        Go back to Step \ref{l17} \;
    }
    {   
        \eIf{$e(\bm{\chi}^{\star}) \geqslant e^{*}$}
        {
            Go back to Step \ref{l17} \;
        }
        {
            \If{$\bm{\chi}^{\star}$ is an integer solution}
            {
                $\bm{\chi}^{*} \leftarrow \bm{\chi}^{\star}$; $e^{*} \leftarrow e(\bm{\chi}^{\star})$ \;
                Go back to Step \ref{l17} \;
            }
        }
        $\textbf{Cut:}$ Search for cutting planes violated by $\bm{\chi}^{\star}$ \;
        \If{any cutting planes are found}
        {
            Add them to the LP relaxation of the problem $\mathcal{Z}_{y}$ and go back to Step \ref{l19} \;
        }
        $\textbf{Branch:}$ Divide $\mathcal{Z}_{y}$ into two subproblems with restricted feasible regions, add them to $\mathcal{Y}$, and go back to Step \ref{l17} \;
    }
}

\end{algorithm}

Through the iterative process of alternating approximations, the feasibility pump algorithm can provide a high-quality initial feasible solution.
We further refine this subcarrier allocation solution using the branch-and-cut method, a proven technique for solving integer programming problems \cite{mitchell2002branch}. 
Specifically, the branch-and-cut method integrates two key components: 1) branch and bound, which iteratively partitions the solution space into subproblems, calculating lower bounds (LBs) and pruning suboptimal regions based on comparisons with the upper bound (UB), and 2) cutting planes, which tighten LP relaxations by introducing additional constraints to exclude non-integral solutions, ensuring convergence to the optimal integer solution efficiently.
% Specifically, the branch-and-cut method consists of two main components: 1) Branch and Bound: 
% It recursively divides the relaxed solution space into smaller subproblems (branching) and computes the lower bound (LB) for each subproblem. 
% Comparing the LB and upper bound (UB) effectively prunes subproblems that cannot contain the optimal solution, thus reducing the search space and narrowing down the solution range. 
% 2) Cutting Planes: During the exploration of the solution space, the cutting plane method generates additional linear inequalities (cutting planes) to tighten the LP relaxations. 
% These inequalities are generated based on the structure of the problem and can exclude non-integral solutions, accelerating the convergence to an integer solution and ensuring the eventual optimal integer solution. 

The latter part of Algorithm \ref{BC_FP}, starting from Step \ref{l15}, outlines the branch-and-cut method. 
It utilizes a tree structure to represent the problem-solving process, where the root node corresponds to the LP relaxation of the original problem, providing an initial LB of the subcarrier energy cost. 
Similarly, the relaxed LP problem is solved by the interior-point method.
The LB is the smallest energy cost from all explored LP relaxations, while the UB is the best energy cost from feasible integer solutions, initialized by the feasibility pump algorithm. 
Through branching, it partitions the feasible regions into smaller subspaces, creating subproblems as child nodes. 
At each node, cutting planes are introduced to  further tighten the solution space, eliminating relaxed solutions that violate the integer constraints.
Pruning is applied to discard branches where the LB exceeds the incumbent UB, reducing unnecessary searches. 
% Fig. \ref{illustration} provides a clear depiction of the workflow of the proposed TSRA algorithm, highlighting the processes of branching, cutting, and pruning.

% \begin{figure}[t]
%     \begin{center}
%         \includegraphics[width=0.375\textwidth]{fig/illustration.pdf}
%         \caption
%         {
%             Illustration of TSRA algorithm using an example with two decision variables. 
%             The black solid circles represent feasible integer solutions, while the blue region depicts the solution space under the LP relaxation.
%             The yellow circle indicates the feasible integer solution obtained from the feasibility pump algorithm, with its corresponding subcarrier energy cost initialized as the UB. 
%             The red solid square marks the incumbent best solution in the LP relaxation. 
%             The purple dashed line represents the added cutting plane.
%         }
%         \label{illustration}
%     \end{center}
% \end{figure} \par

By integrating the branch-and-cut with the feasibility pump, a feasible solution is quickly obtained and iteratively refined to optimality.
The complexity of the proposed TSRA algorithm mainly stems from its second stage, which is the focus of the complexity analysis.
Essentially, the branch-and-cut method establishes a search tree to explore and obtain the optimal solution.
At each node, the LP relaxation of the problem is efficiently solved by the interior-point method.
In the worst case, the time complexity is $\mathcal{O}(2^{|\mathcal{J}|})$ in terms of the number of nodes generated by a brute-force search.
However, in practice, the number of actually explored nodes is generally much smaller due to effective pruning and cutting strategies.
Moreover, the high-quality feasible solution from the first stage provides an initial UB that further facilitates pruning.
Thus, the time complexity of the second stage is $\mathcal{O}(N_{en})$ in terms of the number of actually explored nodes $N_{en}$, where $N_{en} \ll 2^{|\mathcal{J}|}$. 
The space complexity is $\mathcal{O}(N_{an})$, where $N_{an}$ denotes the number of active nodes during the search, including those being processed and those branched out to be explored.
Effective pruning typically ensures that $N_{an} < N_{en}$.

\section{Simulation Results and Discussions}
In this section, we introduce the simulation setup and present the simulation results and discussions of proposed algorithms for two different network load conditions.

\label{SR and D}

\subsection{Simulation Setup}

In our simulation, BS is equipped with $N_{tx} = 16$ transmit antennas, and the user is equipped with $N_{rx} = 4$ receive antennas.
Zero forcing (ZF) equalizer is employed on the user side.
The center frequency is 3.5 GHz and the subcarrier spacing is 15 kHz, with a transmission time interval (TTI) of 1 ms and the numerology $\mu=0$.
In a single TTI or time slot, the resource element (RE) grid for MIMO transmission consists of 2,016 REs, obtained from 144 subcarriers $\times$ 14 symbols, i.e., $K=144$.
For simplicity, $OH$ is fixed at 0.14. 
The transmit power per subcarrier $p$ is set to 1 mW.
In the greedy algorithm, $\eta_{\min}$ is fixed at 0.1.
In TSRA, $N_{\text{flip}}$ is chosen to be 15 to avoid an endless loop and the large constant $\lambda$ is set to 1,000.

Meanwhile, simulations are conducted based on the ray-tracing channel dataset: DeepMIMO \cite{alkhateeb2019deepmimo}.
Using accurate ray-tracing data obtained from Remcom Wireless InSite, clustered delay line (CDL) channels between BS and user are generated according to standard \cite{38.901}.
% DeepMIMO dataset provides candidate BSs and users with various coordinates, with BSs at a height of 6 m and users at 2 m.
In this paper, we focus on using \emph{BS1,2,3} to transmit data to users in \emph{User Grid 1} of DeepMIMO's \emph{O1} scenario, i.e., $M=3$.
A total of 225 users are considered, spanning from \emph{R515} to \emph{R915} in \emph{O1}, arranged in a 15-by-15 grid with 15 users evenly spaced along both the x-axis and the y-axis.
Among them, 10 users are randomly selected in each case to have data rate demands, i.e., $N=10$.
When generating the channel, the maximum number of channel paths is set to 10, though the actual number varies at different locations based on ray-tracing results.

User data demands at each time slot are known from the radio link control (RLC) buffer, calculated as the difference between total arrived and transmitted data. 
Within the relatively short simulation time, data arrivals per TTI for each user are assumed to be stable.
The mean data arrival sizes vary across users to reflect differences in service requirements.
Moreover, we simulate a snapshot of resource scheduling and thus assume that user locations remain unchanged.

\subsection{Comparative Methods}

% \begin{figure}[t]
%     \begin{center}
%         \includegraphics[width=0.44\textwidth]{fig/Locations_of_users_and_BSs.pdf}
%         \caption{Locations of selected users and BSs in DeepMIMO's \emph{O1} scenario.}
%         \label{locations}
%     \end{center}
% \end{figure} \par

We consider the practical implementation of scheduling algorithms under real-time conditions. 
Given the high computation efficiency of NNs and the widespread adoption of RL in addressing resource scheduling problems \cite{he2019joint, chen2024user, zhao2024energy}, the proposed algorithms are benchmarked against RL.
Specifically, the dueling double deep Q-network (D3QN) algorithm is chosen as the RL method to solve the problem.
Compared with DQN, D3QN enhances training stability, improves learning efficiency, and results in more accurate Q-value predictions \cite{wang2016dueling, van2016deep}. 
% In this paper, fully connected NNs are utilized in D3QN, consisting of two hidden layers with 128 neurons each.
% The RL agent undergoes training for 10,000 episodes to ensure convergence.
% To balance exploration and exploitation, we adopt the $\varepsilon$-decay-greedy strategy. 
% Initially, the probability of random action selection $\varepsilon$ is set to 0.99 and decreases by $10^{-4}$ with each learning step until it reaches 0.01. 
% After 4000 exploration steps, the current Q-network is updated every 4 steps, and the target Q-network every 400 steps.
% The minibatch size is 64, and ADAM \cite{kingma2014adam} optimizer with $10^{-4}$ learning rate is used for updating neurons' weights.
% The activation function is rectified linear unit (ReLU).
% The memory buffer size is set to 1,000,000.
State in RL is defined as the concatenation of each user's state, i.e., $\mathfrak{s} = (\mathfrak{s}_1, \ldots, \mathfrak{s}_N)$, where the state of user $n$ is expressed as $\mathfrak{s}_n = (\bm{loc}_n, \vartheta_n, R_{n}^{d}, \hat{R}_{n})$.
Since the problems differ under varying network conditions, the action and reward of RL should be designed accordingly.
In heavy-load network conditions, action $\mathfrak{a}$ is designed to allocate one subcarrier from all DL-BSs to a selected user. 
Action mask is applied to filter available actions where the user's demand is unmet, ensuring the generation of meaningful memories for RL to learn effectively \cite{chen2024user}.
The reward is the increase in the chosen user's weighted demand satisfaction rate, i.e., $\mathfrak{r} = \vartheta_{\mathfrak{a}}$.
The episode terminates when all $K_{\text{remaining}}$ subcarriers are allocated.
In light-load network conditions, action is defined as allocating one subcarrier from a selected combination of cooperating BSs to a user.
The reward is designed to guide the agent in meeting all user demands while minimizing subcarrier energy cost.
If the highest SE TM is selected, user demands can be fully met, but at the cost of higher energy consumption.
Conversely, if the most energy-efficient TM is used to serve users, it may result in unmet user demands after all subcarriers have been allocated.
Therefore, the problem involves a tradeoff between SE and EE.
To this end, we implement two types of RL to solve the problem, namely SE-prioritized and EE-prioritized RL.
The reward is defined as the increase in the weighted sum of SE and EE resulting from the selected action $\mathfrak{a}$. 
It can be expressed as $\mathfrak{r} = \beta_\text{SE} \text{SE}_{\mathfrak{a}} + \beta_\text{EE} \text{EE}_{\mathfrak{a}}$, where $\beta_\text{SE}$ and $\beta_\text{EE}$ represent the importance of SE and EE, respectively.
For SE-prioritized RL, the weights are set as $\beta_\text{SE} = 5$ and $\beta_\text{EE} = 1$. 
For EE-prioritized RL, the weights are set as $\beta_\text{SE} = 1$ and $\beta_\text{EE} = 5$. 
The episode ends when either all user demands are satisfied or all subcarriers are allocated.

\subsection{Evaluation of Proposed Algorithms}

To assess the performance of proposed algorithms under heavy-load and light-load network conditions, we generate independent time slot cases with random user locations, data arrivals, and priority weights.
We randomly select 80\% of the cases for RL's training and use the remaining 20\% for testing.

\subsubsection{Evaluation of Proposed Greedy Algorithm}

\begin{figure}[t]
    \begin{center}
    \subfigure[Running time.]
    {
        % \label{testset_comp}
        \includegraphics[width=0.466\linewidth]{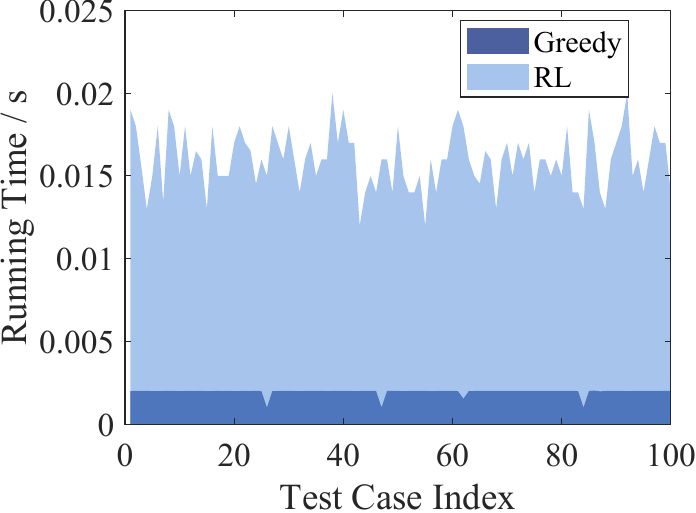}
    }
    \subfigure[CDF of $V$.]
    {
        \includegraphics[width=0.466\linewidth]{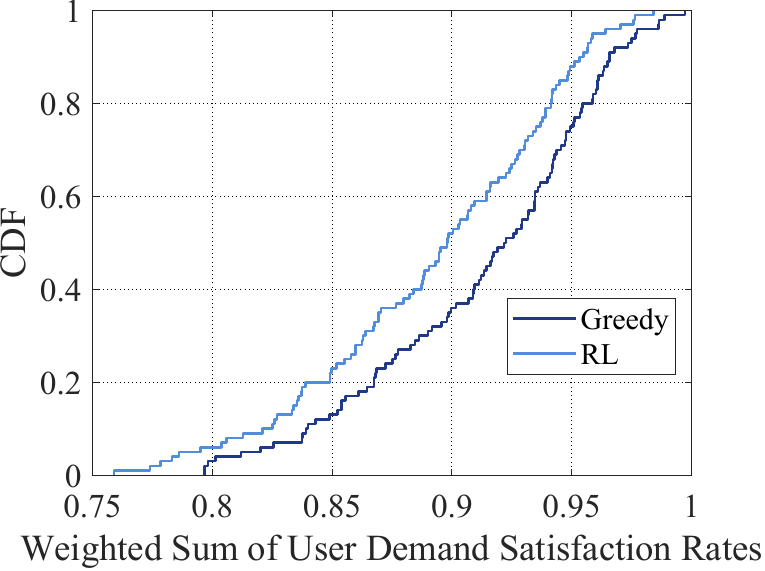}
    }
    \end{center}
    \caption{Comparison between the proposed greedy algorithm and RL under heavy-load network conditions.}
    \label{heavy_load_IS}
\end{figure}

We first compare the proposed greedy algorithm with RL in heavy-load network conditions.
Fig. \ref{heavy_load_IS} illustrates the running time and cumulative density function (CDF) of the weighted sum of user demand satisfaction rates $V$ under heavy-load network conditions.
The greedy algorithm requires less running time and achieves a higher weighted sum of user demand satisfaction rates than RL, as it is a heuristic method and proves to be optimal under heavy-load conditions.

 \par

\subsubsection{Evaluation of Proposed TSRA Algorithm}

\begin{figure}[t]
    \begin{center}
    \subfigure[2 BSs.]
    {
        \includegraphics[width=0.466\linewidth]{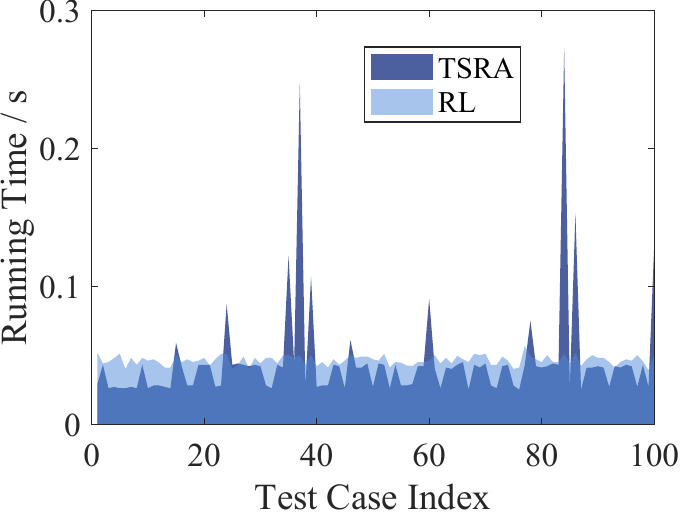}
    }
    \subfigure[3 BSs.]
    {
        \includegraphics[width=0.466\linewidth]{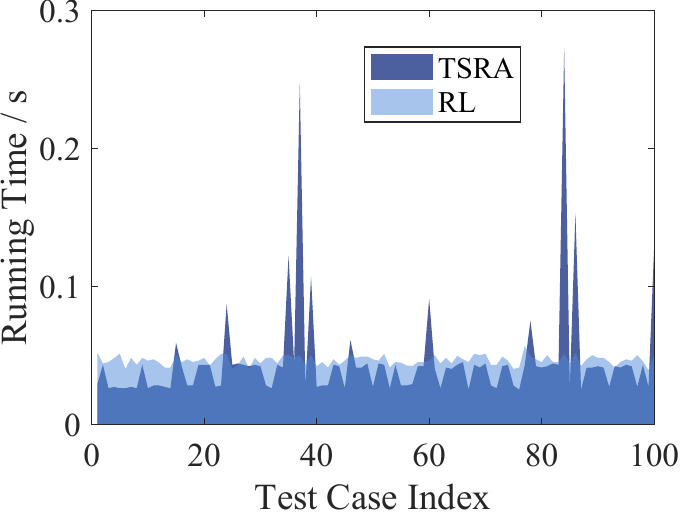}
    }
    \end{center}
    \caption{Comparison of running time between the proposed TSRA algorithm and EE-prioritized RL across networks with different numbers of BSs.}
    \label{rn_light_load}
\end{figure} \par

Given that the second stage of TSRA may require multiple iterations to achieve optimal solutions, we evaluate its and RL's running time for networks consisting of $\emph{BS1,3}$ and $\emph{BS1,2,3}$ in Fig. \ref{rn_light_load}.
Since SE-prioritized and EE-prioritized RL exhibit comparable running times, only the EE-prioritized RL is considered in this comparison.
The running time per episode for RL is relatively stable, whereas the TSRA algorithm may require additional computational time to iteratively reach optimal solutions, especially in more complex scenarios of three BSs.
To ensure a fair comparison, we impose a time limit on the TSRA algorithm. 
Using RL's average running time as the baseline, the TSRA algorithm outputs the best solution found within the same runtime, denoted as “TSRA with time limits”.

\begin{figure}[t]
    \begin{center}
    \subfigure[Success rate of problem solving.]
    {
        \includegraphics[width=0.466\linewidth]{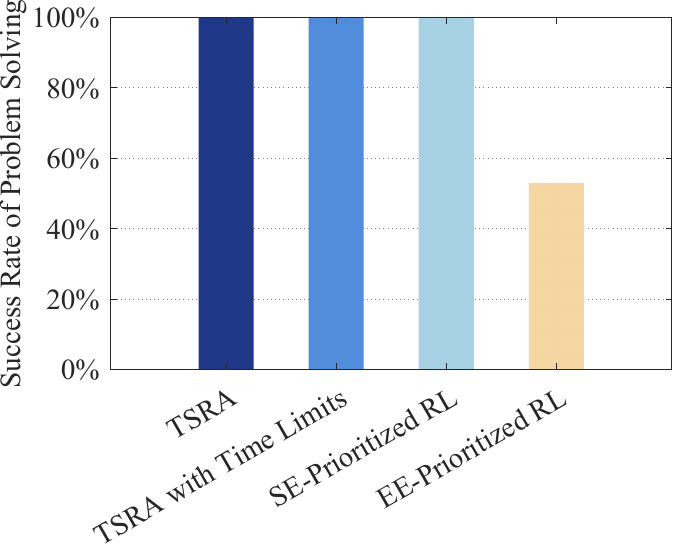}
    }
    \subfigure[KDE of $e$.]
    {
        \includegraphics[width=0.466\linewidth]{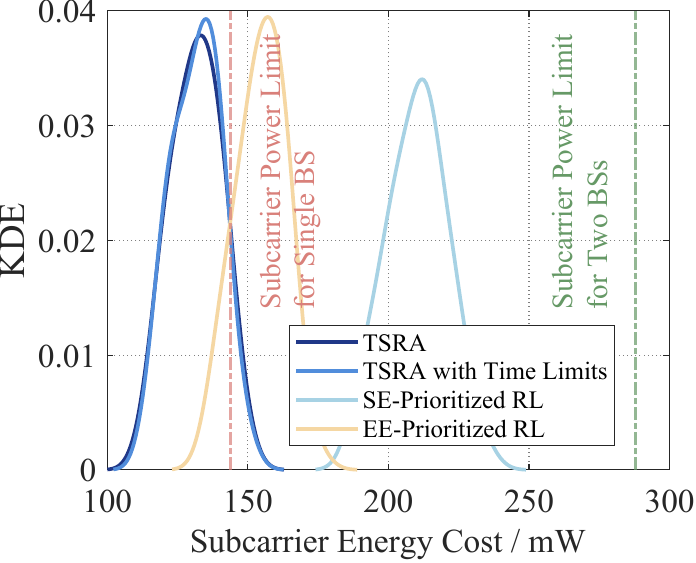}
    }
    \end{center}
    \caption{Comparison between the proposed TSRA algorithm and RL when the network consists of 2 BSs.}
    \label{c_light_load_2BS_IS}
\end{figure} \par

Since RL cannot guarantee to meet the constraint of satisfying all user demands, we compare the TSRA algorithm and RL in terms of the percentage of successful problem-solving cases and plot the kernel density estimation (KDE) of energy cost $e$ for all algorithms in their common successful cases.
KDE is an estimation of the probability density function (PDF) of a variable using a Gaussian kernel.
Fig. \ref{c_light_load_2BS_IS} depicts the scenario involving $\emph{BS1,3}$.
Although TSRA is subject to a running time limit, it still quickly obtains a feasible solution through the feasibility pump algorithm, successfully solving the problem. 
As expected, SE-prioritized RL outperforms EE-prioritized RL in terms of problem-solving success rate, while EE-prioritized RL achieves lower energy cost. 
The TSRA algorithm with time constraints effectively approximates the optimal solution and outperforms both RL methods, demonstrating its effectiveness.

\begin{figure}[t]
    \begin{center}
    \subfigure[Success rate of problem solving.]
    {
        \includegraphics[width=0.466\linewidth]{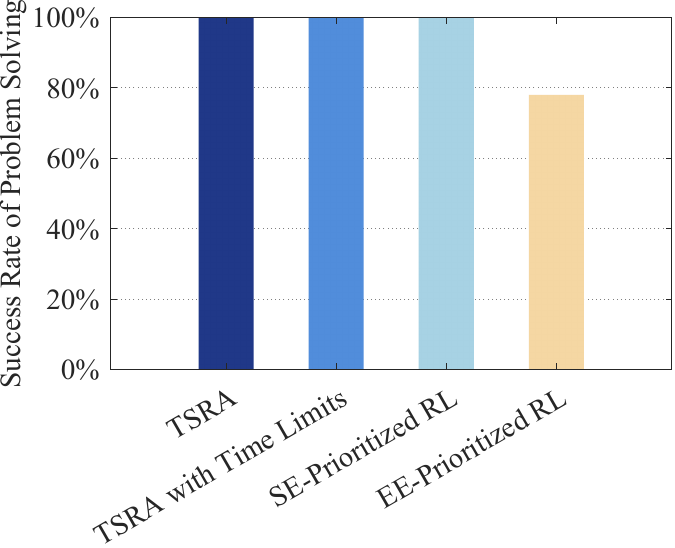}
    }
    \subfigure[KDE of $e$.]
    {
        \includegraphics[width=0.466\linewidth]{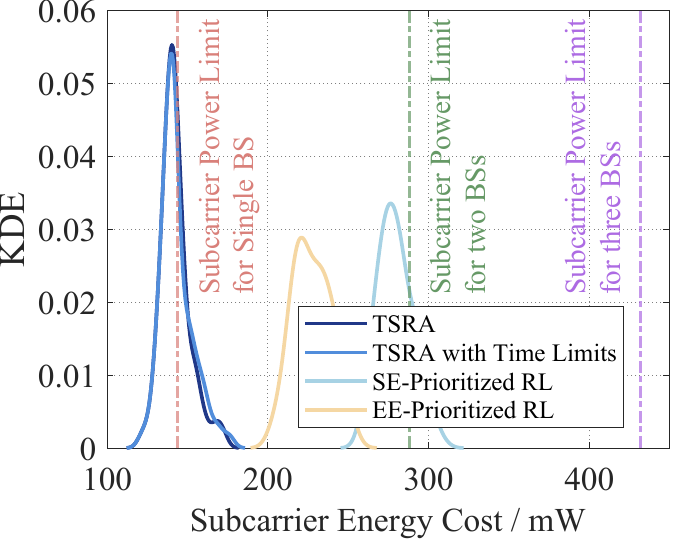}
    }
    \end{center}
    \caption{Comparison between the proposed TSRA algorithm and RL when the network consists of 3 BSs.}
    \label{c_light_load_3BS_IS}
\end{figure} \par

Compared to the two-BS scenario, three-BS cooperation expands the action space, making it more challenging for RL to explore and learn effectively.
As illustrated in Fig. \ref{c_light_load_3BS_IS}, although EE-prioritized RL improves the success rate of problem-solving, it tends to favor JT to meet user demands, resulting in higher energy cost.
% For TSRA with time constraints, although the gap from the optimal solution slightly widens in the three-BS scenario, it still achieves competitive suboptimal solutions.
For TSRA under time constraints, it still yields competitive near-optimal solutions.
Meanwhile, the subcarrier energy consumed by TSRA is significantly lower than the system's maximum energy consumption, demonstrating the necessity and value of energy-efficient scheduling.
% the subcarrier power limits demonstrate that the TSRA algorithm effectively minimizes the energy cost.
% while fully meeting all user demands.

\subsection{Evaluation of Feedback-Free Resource Scheduling}

We further compare resource allocation in feedback-free and feedback-based systems, so as to validate the advantage of flexible cooperation in FD-RAN.
The greedy and time-constrained TSRA algorithms demonstrate superior overall performance compared to RL, thus, they are chosen as the resource scheduling algorithms for heavy-load and light-load network conditions in subsequent comparisons.
To implement feedback-free
resource scheduling in FD-RAN, corresponding transmission parameters at each location are pre-calculated using historical channel data via variational autoencoder (VAE) \cite{liu2025enabling}.
To implement feedback-based resource scheduling, both implicit and explicit CSI  feedback schemes are considered.
The former reports a precoding matrix indicator (PMI) from a predefined codebook, while the latter reports the optimal precoder.
Accordingly, these are referred to as “feedback PMI” and “feedback optimal precoder”, respectively.
Considering the increasing overheads of pilots and feedback, as well as the computational complexity of joint transmission parameters for multi-BS cooperation under practical environment, feedback-based JT is restricted to two BSs, referred to as “CoMP”.
Without loss of generality, single BS transmission and CoMP in this paper utilize $\emph{BS1}$ and $\emph{BS1,3}$, respectively.

Moreover, we perform evaluation under independent and  consecutive scheduling decisions.
To simulate time-varying channels, distinct channel samples are generated for each time slot, yielding a channel coherence time of 1 ms. While the path loss is assumed to remain constant at a specific location, different random seeds are set to reflect small-scale environmental variations, incorporating ray-tracing results into the CDL model in the static \emph{O1} scenario.
In independent scheduling, each test case contains two time slots with two independent
channel samples, assuming no delay in executing algorithms and activating decisions (i.e., $d_2 = d_3 = 0$).
For feedback-based schemes, CSI feedback is obtained from either the first or second channel sample to simulate a 1 ms (1 TTI) feedback delay or no delay, referred to as “with delay” and “without delay”, respectively.
The scheduling decision is made for the second time slot, assuming no data accumulation.
In consecutive scheduling, each test case comprises 100 consecutive time slots spanning 100 ms, with 100 distinct channel samples.
Untransmitted data from the previous time slots are queued in the RLC buffer for subsequent transmission.
This setup enables a performance comparison under more realistic conditions, incorporating various delays configured as $d_1 = 2$ and $d_2 = d_3 = 3$.
The location/CSI reporting period and scheduling period are set to $T_{rp} = 3$ and $T_{sc} = 10$, respectively.
Note that both delays and periods are represented in units of time slots.
User locations, data arrivals, and priority weights are randomly initialized in each test case for both independent and consecutive scheduling decisions.

\begin{figure}[t]
    \begin{center}
        \includegraphics[width=0.45\textwidth]{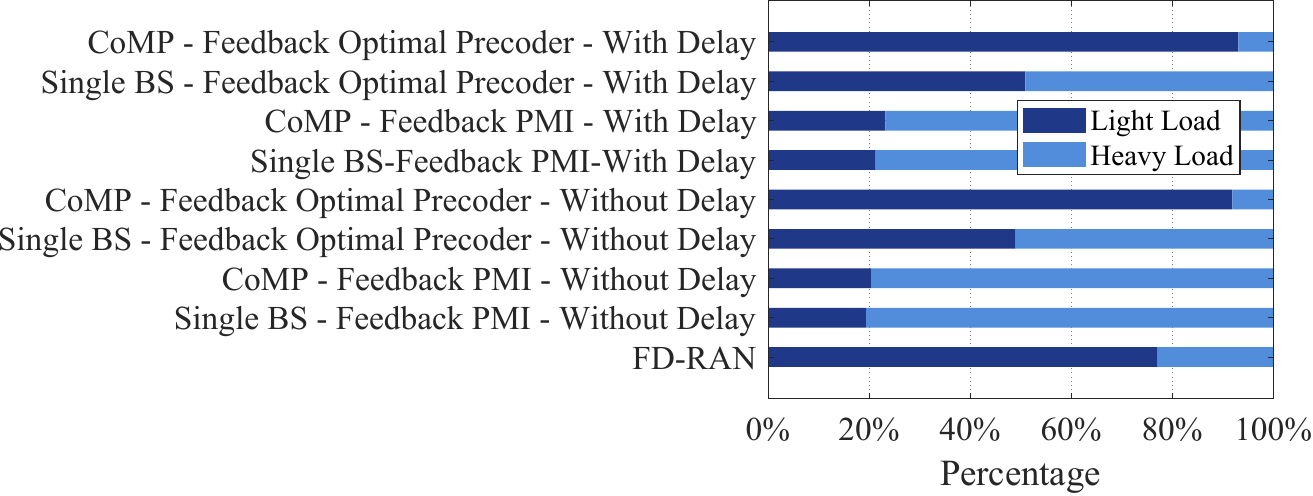}
        \caption{Proportion of heavy-load and light-load network conditions across different transmission schemes under independent scheduling decisions.}
        \label{heavy_light_ratio_IS}
    \end{center}
\end{figure} \par

\subsubsection{Comparison Under Independent Scheduling Decisions}
We first compare feedback-free and feedback-based resource scheduling under independent scheduling decisions.
Fig. \ref{heavy_light_ratio_IS} presents the proportion of heavy-load and light-load network conditions for various transmission schemes, reflecting the diversity of the test cases.
In independent scheduling, where cumulative data in the buffer is not considered, the classification of the network condition as heavy-load or light-load depends on the data rate provided by the specific transmission scheme. 
The proportion of light-load network conditions under FD-RAN is relatively high, approaching that of CoMP with optimal precoder feedback, highlighting its strong scheduling capability in effectively meeting user demands.

\begin{figure}[t]
    \begin{center}
    \subfigure[Without delay.]
    {
        \label{CDF_V_IS_without_delay}
        \includegraphics[width=0.466\linewidth]{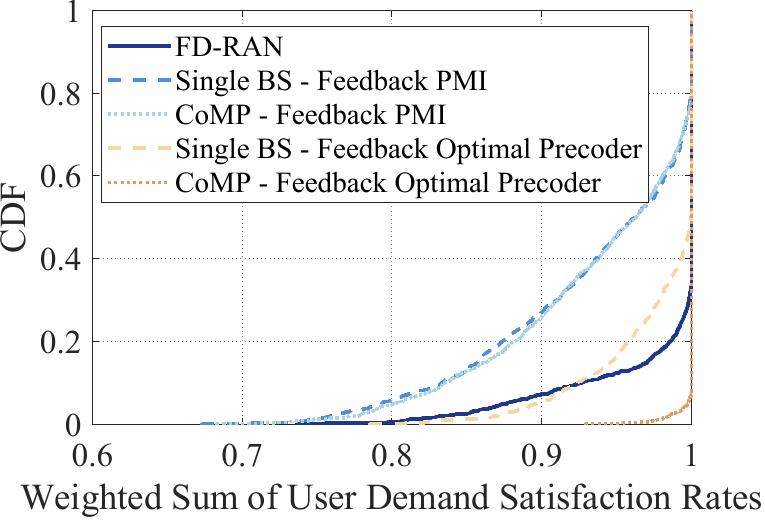}
    }
    \subfigure[With delay.]
    {
        \label{CDF_V_IS_with_delay}
        \includegraphics[width=0.466\linewidth]{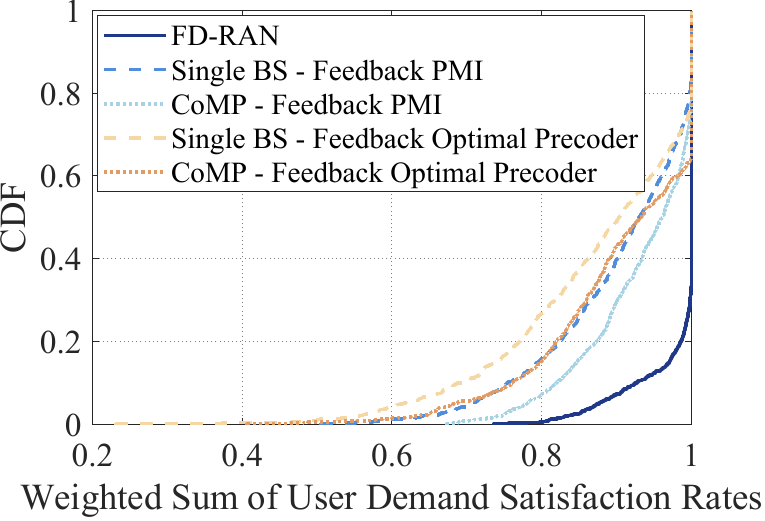}
    }
    \end{center}
    \caption{Comparison of $V$ between feedback-free and feedback-based resource scheduling under independent scheduling decisions.}
    \label{CDF_V_IS}
\end{figure} \par

Note that $V$ should equal 1 in light-load conditions where all user demands are fully satisfied. 
Therefore, the CDF of $V$ can be used to compare the overall scheduling performance, illustrated in Fig. \ref{CDF_V_IS}.
In independent scheduling, $V$ is defined as the ratio of correctly transmitted data to the data arriving in the same time slot.
Under ideal conditions without feedback delay, optimal precoders yield more accurate directional beams than codebooks, thereby achieving superior performance, as illustrated in Fig. \ref{CDF_V_IS_without_delay}.
The feedback-free transmission adopted by FD-RAN exploits historical channel data to determine fixed transmission parameters for each location. 
While this may lead to lower data rates than feedback-based schemes without delays, it enhances robustness in the presence of delays.
As shown in Fig. \ref{CDF_V_IS_with_delay}, such robustness allows FD-RAN to outperform other approaches when feedback delays are present.
In contrast, employing optimal precoders suffers greater performance degradation than PMI when outdated CSI is used.

\begin{figure}[t]
    \begin{center}
    \subfigure[CDF of $e$.]
    {
        \label{CDF_e_light_load_IS}
        \includegraphics[width=0.466\linewidth]{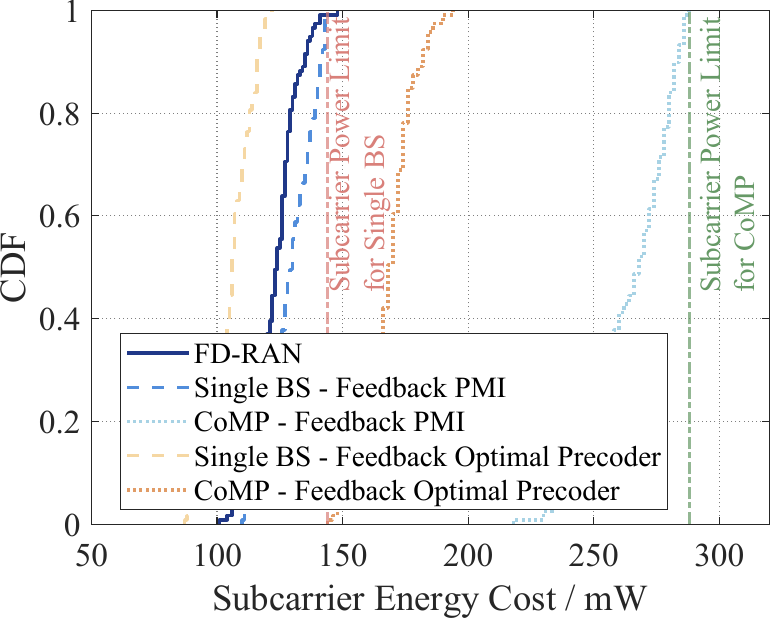}
    }
    \subfigure[$V$.]
    {
        \label{V_light_load_IS}
        \includegraphics[width=0.466\linewidth]{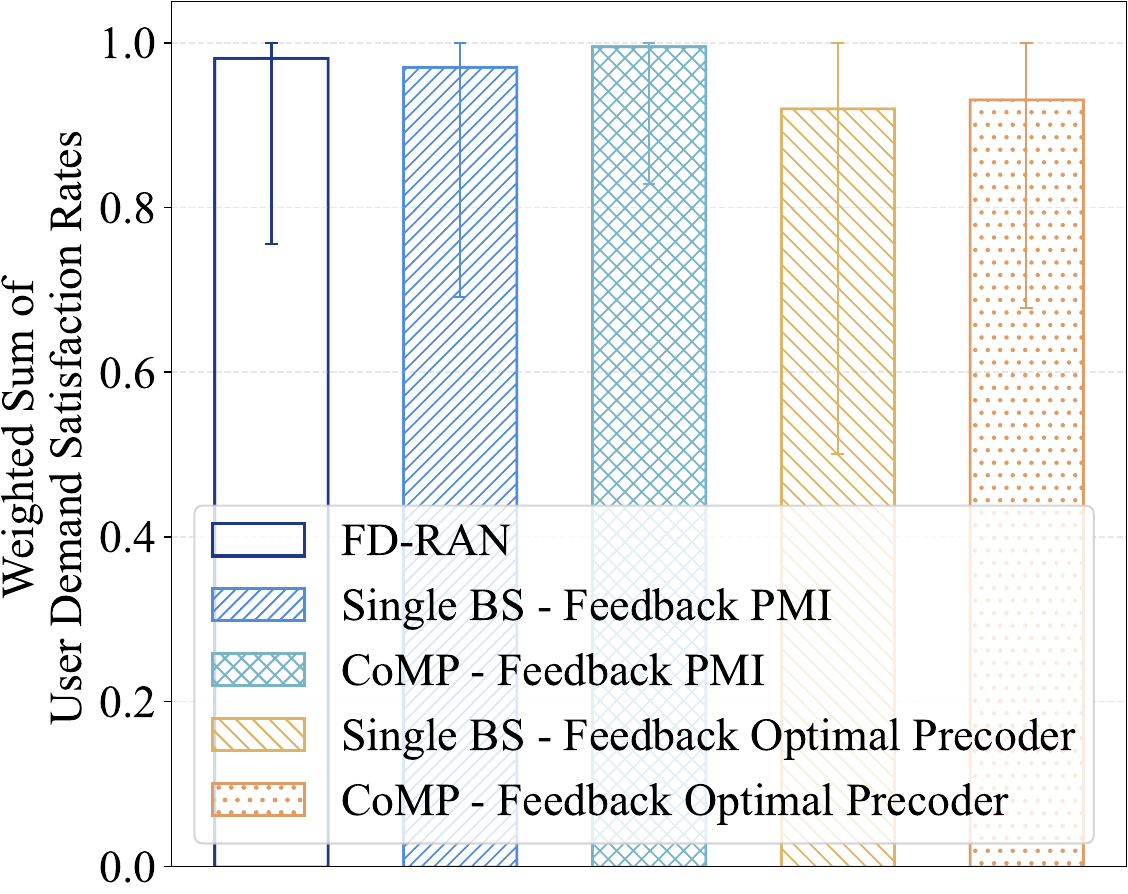}
    }
    \end{center}
    \caption{Comparison between feedback-free and feedback-based resource scheduling with feedback delay under light-load network conditions.}
    \label{c_light_load_IS}
\end{figure} \par

When delving into the test data, we find some test cases fall under light-load conditions for all transmission schemes.
Thus, their EE performance is evaluated in the presence of feedback delay, as shown in Fig. \ref{c_light_load_IS}.
It is observed that FD-RAN, enabled by flexible cooperation, can dynamically select the most appropriate BS for each user, thereby achieving lower subcarrier energy cost than single-BS transmission with PMI feedback in most cases.
Although explicit feedback with optimal precoders theoretically enables higher data rates, the high beam precision renders it more sensitive to feedback delays, resulting in greater rate fluctuations and unmet user demands.
This is reflected in both the mean and variations of $V$ in Fig. \ref{V_light_load_IS}.
In contrast, FD-RAN achieves comparable user demand satisfaction rates to PMI-based CoMP with lower subcarrier energy cost, due to its insensitivity to delay and flexibility in selecting the most energy-efficient single-BS transmission.

\subsubsection{Comparison Under Consecutive Scheduling Decisions}
For consecutive scheduling decisions, we consider both low and high data arrival rates at user buffers to simulate different network load conditions.

\begin{figure}[t]
    \begin{center}
    \subfigure[Arrived, transmitted and remaining data in the buffer.]
    {
        \includegraphics[width=0.75\linewidth]{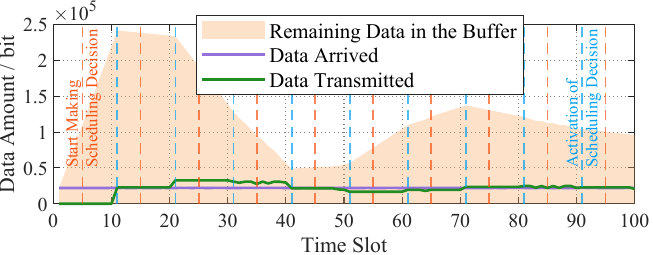}
    } \\
    \subfigure[Subcarrier energy cost.]
    {
        \includegraphics[width=0.8\linewidth]{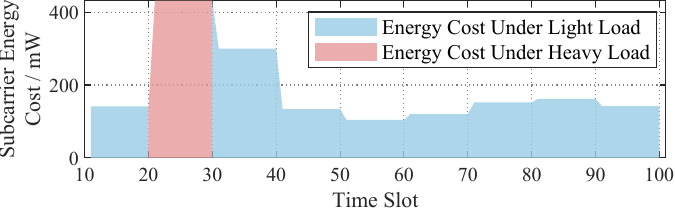}
    }
    \end{center}
    \caption{Visualization of trends of variables in FD-RAN during data transmission from the perspective of all users.}
    \label{illustration_whole}
\end{figure} \par

\begin{figure}[t]
    \begin{center}
        \includegraphics[width=0.42\textwidth]{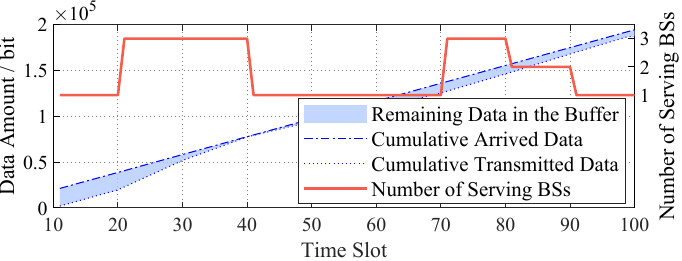}
        \caption{Visualization of trends of variables in FD-RAN during data transmission from the perspective of one user.}
        \label{illustration_SU}
    \end{center}
\end{figure} \par

\begin{figure}[t]
    \begin{center}
        \includegraphics[width=0.4\textwidth]{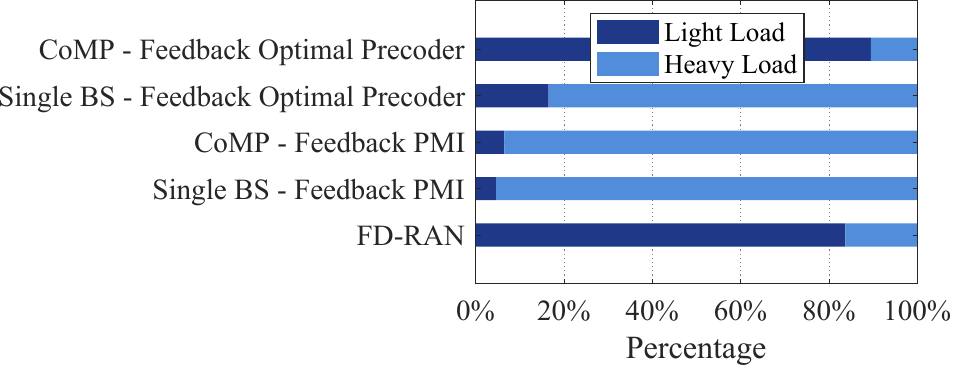}
        \caption{Proportion of heavy-load and light-load network conditions across different transmission schemes under consecutive scheduling decisions.}
        \label{heavy_light_ratio_CS}
    \end{center}
\end{figure} \par

\begin{figure}[t]
    \begin{center}
    \subfigure[Average $V$.]
    {
        \label{comp_eta_low_demand_CS}
        \includegraphics[width=0.466\linewidth]{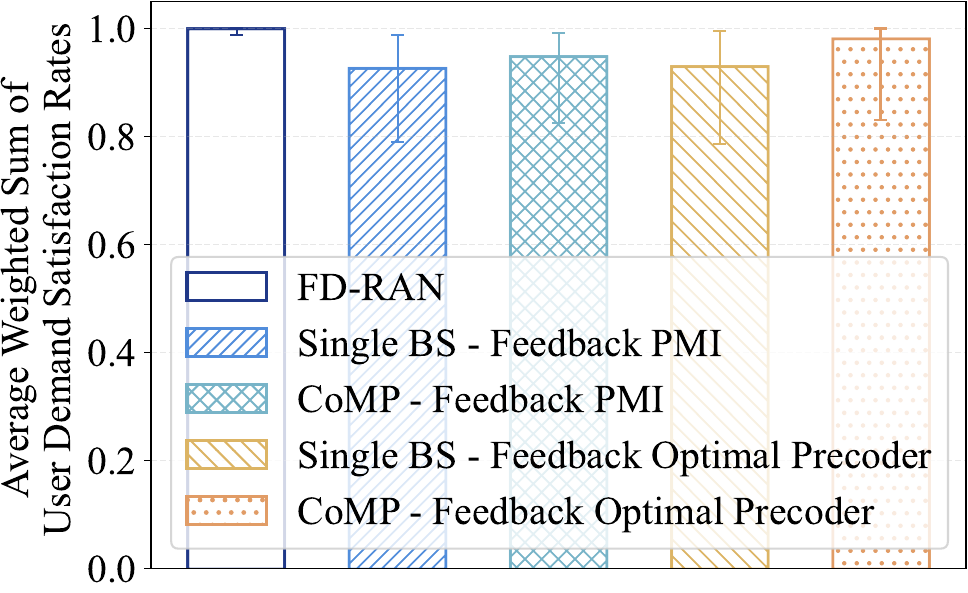}
    }
    \subfigure[Bad CQI ratio.]
    {
        \label{comp_0ratio_low_demand_CS}
        \includegraphics[width=0.466\linewidth]{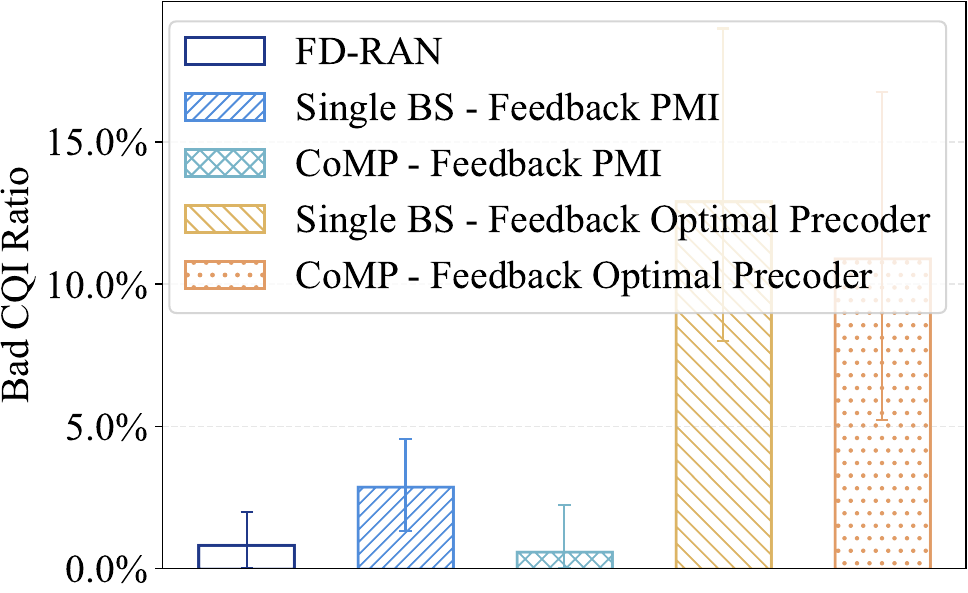}
    }
    \end{center}
    \caption{Comparison between feedback-free and feedback-based scheduling with a low mean data arrival rate under consecutive scheduling decisions.}
\end{figure} \par

\begin{figure}[t]
    \begin{center}
        \includegraphics[width=0.43\textwidth]{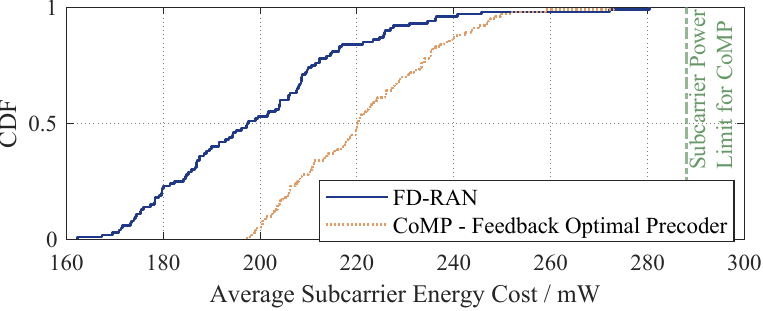}
        \caption{Comparison of $e$ between FD-RAN and CoMP with a low mean data arrival rate under consecutive scheduling decisions.}
        \label{comp_CDF_energy_low_demand_CS}
    \end{center}
\end{figure} \par

\begin{figure}[t]
    \begin{center}
    \subfigure[Average $V$.]
    {
        \includegraphics[width=0.466\linewidth]{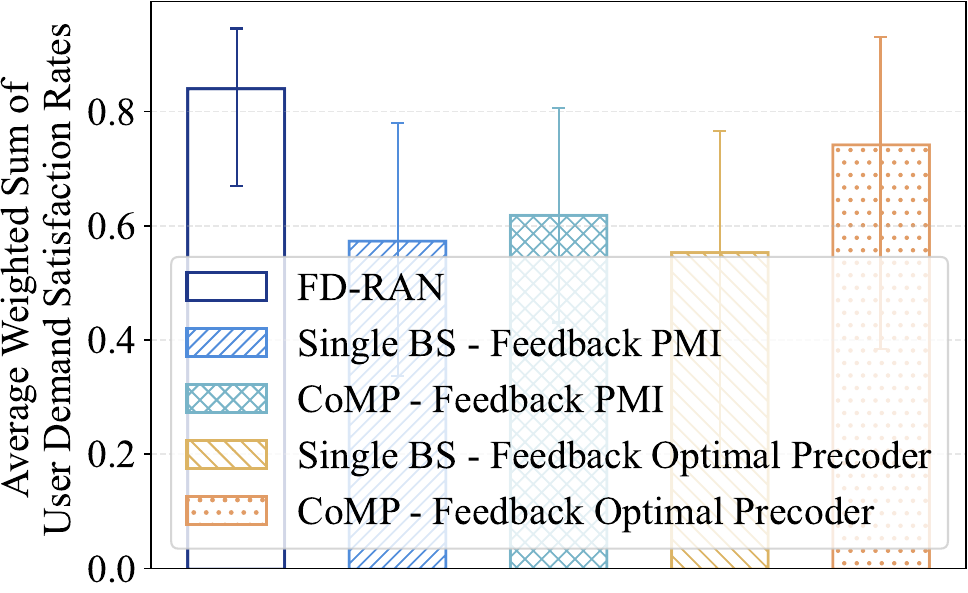}
    }
    \subfigure[Bad CQI ratio.]
    {
        \includegraphics[width=0.466\linewidth]{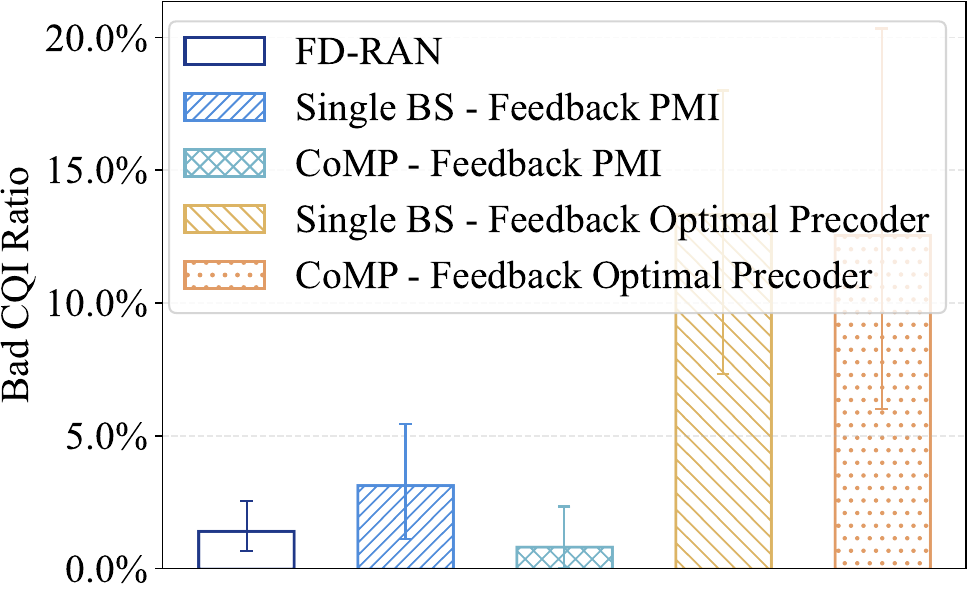}
    }
    \end{center}
    \caption{Comparison between feedback-free and feedback-based scheduling with a high mean data arrival rate under consecutive scheduling decisions.}
    \label{comp_high_demand_CS}
\end{figure} \par

% \subsubsection{Low Mean Data Arrival Rate}
First, we compare feedback-free and feedback-based resource allocation under a low mean data arrival rate in different users' buffers.
Fig. \ref{illustration_whole} and Fig. \ref{illustration_SU} illustrate the trends of variables during the data transmission when the transmission scheme is FD-RAN, from the perspectives of all users and one user in the network, respectively.
The former figure presents the network's transitions between heavy-load and light-load conditions over consecutive time slots, as well as the fluctuation of the total data amount in user buffers.
The latter highlights FD-RAN's flexibility in selecting serving BSs to meet user demands.
In consecutive scheduling decisions, 
the network may transition to the heavy-load condition as data accumulates in the buffer when the selected transmission scheme provides insufficient data rates.
Fig. \ref{heavy_light_ratio_CS} visualizes the proportion of heavy-load and light-load network conditions for various transmission schemes under consecutive scheduling decisions.
CoMP with optimal precoder feedback and FD-RAN with flexible cooperation can provide higher data rates, helping to maintain the network in light-load conditions.

We also plot the average $V$ for various transmission schemes, as shown in Fig. \ref{comp_eta_low_demand_CS}.
In consecutive scheduling, $V$ is defined as the ratio of the total transmitted data during the scheduling period to the cumulative data preceding it.
FD-RAN achieves the highest average $V$, reaching 1 in most cases.
This is because the feedback-free transmission provides stable data rates with representative transmission parameters, resulting in fewer occurrences of bad CQI, as shown in Fig. \ref{comp_0ratio_low_demand_CS}. 
Bad CQI refers to the selection of a higher CQI that results in zero throughput and causes data to accumulate in the buffer.
Since FD-RAN and CoMP with optimal precoder feedback are more effective in maintaining the network in light-load conditions, their subcarrier energy costs are compared in Fig. \ref{comp_CDF_energy_low_demand_CS}. 
Due to the flexibility in selecting the most suitable cooperation set, FD-RAN demonstrates lower energy costs.

When the mean data arrival rate at user buffers is high, total user demands exceed the network’s capacity, making it impossible to fully satisfy them within each scheduling period. 
As a result, the network remains in heavy-load conditions.
Nevertheless, FD-RAN still delivers the highest average $V$ by leveraging cooperation among all potential BSs while maintaining relatively few instances of bad CQI due to appropriate fixed transmission parameters, as shown in Fig. \ref{comp_high_demand_CS}.
Regardless of whether the mean data arrival rate is low or high, FD-RAN demonstrates consistently reliable performance through flexible cooperation, efficiently utilizing resources to meet diverse user demands.

\section{Conclusion}
\label{Con}
In this paper, we have studied how to realize feedback-free DL resource scheduling to enable multi-BS flexible cooperation in FD-RAN.
A problem of joint cooperation set selection and subcarrier allocation is formulated, aiming primarily to satisfy personalized user demands.
Then, the problem is decomposed based on network load conditions.
In heavy-load conditions, priorities of user services and fairness are further considered, where a greedy algorithm is proposed and proves optimal.
In light-load conditions, the problem is modeled as reducing subcarrier energy cost while still meeting user demands, where a TSRA algorithm is proposed to obtain a near-optimal solution within a short time.
Simulation results based on ray-tracing channel data demonstrate that the proposed algorithms achieve higher resource utilization efficiency than RL algorithms under the same runtime. 
Moreover, the feedback-free resource scheduling approach outperforms the two-BS CoMP scheme.
Our future work will explore more diverse and advanced transmission methods such as NOMA and multi-user MIMO to further enhance the resource utilization flexibility of FD-RAN.

\ifCLASSOPTIONcaptionsoff
  \newpage
\fi

\bibliographystyle{IEEEtran}
\bibliography{IEEEarbv,ref}

\end{document}